\documentclass[aps,preprint,showpacs,superscriptaddress,groupedaddress]{revtex4}  %
\usepackage{graphicx}  %
\usepackage{dcolumn}   %
\usepackage{bm}        %
\usepackage{amssymb}   %
\usepackage{nicefrac}
\usepackage{mathrsfs}
\usepackage{slashed} 
\usepackage{array} 
\usepackage{epsfig} 
\newcommand{\bra}[1]{\big\langle \, #1\,\big\vert}
\newcommand{\ket}[1]{\big\vert\, #1\,\big\rangle}
\newcommand{\bracket}[2]{\big\langle \, #1 \big\vert \, #2\,\big\rangle}
\newcommand{\tg}{\mathop{\mathrm{tg}}}
\newcommand{\ch}{\mathop{\mathrm{ch}}}
\newcommand{\sh}{\mathop{\mathrm{sh}}}

\begin{document}

\title{Experimental test of the time-dependent Wigner inequalities for
  neutral pseudoscalar meson systems}

\author{N. Nikitin}
\affiliation{Lomonosov Moscow State University Department of Physics, Russia}
\affiliation{Lomonosov Moscow State University Skobeltsyn Institute of
  Nuclear Physics, Russia}
\affiliation{Institute for Theoretical and Experimental Physics, Russia}
\author{V. Sotnikov}
\affiliation{Lomonosov Moscow State University Department of Physics, Russia}
\affiliation{Lomonosov Moscow State University Skobeltsyn Institute of
  Nuclear Physics, Russia}
\author{K. Toms}
\affiliation{Department of Physics and Astronomy, University of New
  Mexico, USA}
\date{\today}

\begin{abstract}
Recently a new class of time-dependent Bell
inequalities in Wigner form was introduced. The structure of the
inequalities allows experimental studies of quantum and open
quantum systems in external fields. In this paper we study the
properties of the time-dependent Wigner inequalities using the time
evolution of neutral pseudoscalar mesons. It is shown that it is
always possible to find a range of parameters to test for 
violation in an experimentally accessible area. The effect of the
relaxation of the inequalities for large time scales is
demonstrated.
\end{abstract}

\maketitle

\section{Introduction}
In \cite{Nikitin:2014yaa} a new class of time-dependent Bell
inequalities in Wigner form was introduced. The structure of the
inequalities allows experimental studies of the quantum and open
quantum systems in an external fields. In this paper we study the
properties of the time-dependent Wigner inequalities using the time
evolution of neutral pseudoscalar mesons.

In \cite{Einstein:1935rr} the following question was raised for the
first time: can the properties of a macro-system which in
quantum theory are described by noncommuting operators be
simultaneously the elements of reality (i.e. to exist simultaneously),
even if these properties can not be measured by any macro-device? %
According to the Copenhagen interpretation of quantum
mechanics, the answer is no \cite{Bohr:1935af}. Bohr demonstrated
the fundamental difference between the statistical and the Copenhagen
interpetations, but did not give a conclusive proof
\cite{Einstein:1935rr}. An attempt to move the problem of the
simultaneous existence of the elements of physical reality from
the gedanken to the experimental realm has been made by J.~Bell
\cite{Bell:1964kc,Bell:1964fg,nature510}. Bell's idea has been further
developed by  Clauser, Horne, Shimony, and
Holt \cite{Clauser:1969ny}. Since then the idea has been thoroughly
studied, developed, and criticized by many Bell's proponents and
opponents
\cite{epr-books,Reid:2009zz,Rosset:2014tsa,RevModPhys.86.419}.

We address here the question of how to express the fact, that some set
of a micro-system characteristics (observables) is
simultaneously the set of elements
of the physical reality, even if that set can not be simultaneously
measured by any macro-device. One possibility is to assert
that the joint probability of simultaneous existence of members of the
set is non-negative. In quantum mechanics the
probability is the universal object. %
Such a proposition (however without the clear statement) was used by Bell in
\cite{Bell:1964kc}. The idea of the non-negativity of the
joint probability was proposed by E. Wigner \cite{wigner}.

Following Wigner's approach, let us suppose that a quantum system
decays at the time $t_0$ into two subsystems ``1'' and ``2'', each
having three observables $a$, $b$, and $c$. Let each observable be
dichotomic variable (able to have only two distinct values). For
simplification let us set these values to $\pm 1$. We will use the
following definitions: if the observable $a$ is equal to $+1$ we
denote this as $a^{(1)}_+$, and so on. At the time $t_0$ let all
three observables to satisfy the anticorrelation condition 
\begin{eqnarray}
\label{pm=mp2}
n_{\pm}^{(1)}(t_0)\, =\, -\, n_{\mp}^{(2)}(t_0),
\end{eqnarray}
where $n^{(i)} = \{a^{(i)}, \, b^{(i)},\, c^{(i)} \}$ and $i =
\{1,\,2\}$.

An example of such a system is a pseudoscalar
particle which decays into a fermion-antifermion pair with a Hamiltonian
\begin{eqnarray}
\label{Heff_for_PS2ff}
\mathcal{H}^{(PS)}(x)\, =\, g\,\varphi (x)\,\left (\bar f(x)\,\gamma^5\, f(x)\right )_N, 
\end{eqnarray}
which automatically provides full anticorrelation of the fermions' spin
projections onto any direction.  In (\ref{Heff_for_PS2ff}), the $\varphi
(x)$  is a pseudoscalar field, and $\bar f(x)$ and $f(x)$ are fermionic fields.

Let $a$, $b$, and $c$ exist simultaneously as the elements of
physical reality, i.e. any of their double and triple joint
probabilities are non-negative. Then, under the assumption of locality
at time $t_0$ and using Kolmogorov's axiomatics of probability
theory, the following inequality can be obtained \cite{Nikitin:2014yaa}:

\begin{eqnarray}
\label{W-B-3}
w\left (a^{(2)}_+, b^{(1)}_+, t_0 \right )\,\le\, 
w\left (c^{(2)}_+, b^{(1)}_+, t_0 \right )\, +\,
w\left (a^{(2)}_+, c^{(1)}_+, t_0 \right ). 
\end{eqnarray}
If one drops the $t_0$ from (\ref{W-B-3}), it transforms into the Wigner
inequality \cite{wigner}
\begin{eqnarray}
\label{W-B-2}
w\left (a^{(2)}_+, b^{(1)}_+\right )\,\le\, 
w\left (c^{(2)}_+, b^{(1)}_+\right )\, +\,
w\left (a^{(2)}_+, c^{(1)}_+\right ). 
\end{eqnarray}

We name the inequalities (\ref{W-B-3}) and
(\ref{W-B-2}) time-independent or static Wigner inequalities, 
to distinguish them from the time-dependent inequalities in 
\cite{Nikitin:2014yaa}. The observables $a$, $b$, and $c$
may correspond to non-commuting operators and, hence, cannot be
simultaneously measured by any macro-device. As was mentioned in
\cite{Nikitin:2009sr}, the Wigner inequalities are more suitable for
experimental tests, due to the fact that the probabilities, unlike
the correlators, are well defined in both non-relativistic quantum
theory and in quantum field theory.

In \cite{Nikitin:2014yaa},  a new class of Wigner inequalities was
obtained with a direct dependence on time: 
\begin{eqnarray}
\label{W-B-4}
&& w\left (a^{(2)}_+, b^{(1)}_+, t \right )\,\le \\
&\le&
w\left (a^{(2)}_+(t_0) \to a^{(2)}_+ (t) \right )\,
\left (
w\left (b^{(1)}_+(t_0) \to b^{(1)}_+ (t) \right )\, +\, 
w\left (b^{(1)}_-(t_0) \to b^{(1)}_+ (t) \right )
\right )\, 
w\left (a^{(2)}_+, c^{(1)}_+, t_0 \right )\, +\nonumber\\
&+&
w\left (a^{(2)}_-(t_0) \to a^{(2)}_+ (t) \right )\,
\left (
w\left (b^{(1)}_+(t_0) \to b^{(1)}_+ (t) \right )\, +\, 
w\left (b^{(1)}_-(t_0) \to b^{(1)}_+ (t) \right )
\right )\, 
w\left (a^{(2)}_-, c^{(1)}_+, t_0 \right )\, +\nonumber\\
&+&
w\left (b^{(1)}_+(t_0) \to b^{(1)}_+ (t) \right )\,
\left (
w\left (a^{(2)}_+(t_0) \to a^{(2)}_+ (t) \right )\, +\, 
w\left (a^{(2)}_-(t_0) \to a^{(2)}_+ (t) \right )
\right )\,
w\left (c^{(2)}_+, b^{(1)}_+, t_0 \right )\, + \nonumber\\ 
&+&
w\left (b^{(1)}_-(t_0) \to b^{(1)}_+ (t) \right )\,
\left (
w\left (a^{(2)}_+(t_0) \to a^{(2)}_+ (t) \right )\, +\, 
w\left (a^{(2)}_-(t_0) \to a^{(2)}_+ (t) \right )
\right )\,
w\left (c^{(2)}_+, b^{(1)}_-, t_0 \right ).\nonumber 
\end{eqnarray}
For closed quantum systems, $w\left (a^{(2)}_-(t_0) \to a^{(2)}_+ (t)
\right ) =  w\left (b^{(1)}_-(t_0) \to b^{(1)}_+ (t) \right ) = 0$,
while $w\left (a^{(2)}_+(t_0) \to a^{(2)}_+ (t) \right ) = w\left
  (b^{(1)}_+(t_0) \to b^{(1)}_+ (t) \right ) = 1$. Hence 
(\ref{W-B-4}) reduces to (\ref{W-B-3}), as it should be from physical
point of view. The inequality (\ref{W-B-3}), in turn, is equivalent to
the time-independent inequality (\ref{W-B-2}). 

The Leggett-Garg inequalities are based on the
idea of macroscopic realism \cite{Leggett:1985zz}. They resemble Bell
inequalities \cite{Clauser:1969ny} but, instead of the simultaneous
correlation of the two observables, they concern the correlation
between the values of a single observable at different points of 
time. In experiment they are closely related to the weak
(non-invasive) measurements \cite{Aharonov:1988xu,korotkov-2006},
involving, for example, nanomechanical resonators \cite{NatureComm-2011}
or discrete lattices \cite{ideal-2015}.

The main distinction between the Leggett-Garg inequalities and
(\ref{W-B-4}) is the fact that the test of (\ref{W-B-4}) does not
require weak measurements. The measurement is fully invasive. That
opens the experimental possibility to verify (\ref{W-B-4}) at
contemporary high energy physics detectors like LHCb, ATLAS, CMS, and
Belle II.

We study here the violation of the time-dependent
inequality (\ref{W-B-4}) in systems of neutral pseudoscalar
mesons. Many papers attempt to include 
time-dependence into the static Wigner inequalities
\cite{Bell:1964kc,Clauser:1969ny,wigner}, and studies of
the obtained time-dependent inequalities in quantum theory are
available \cite{Privitera:1992xc}
-- \cite{Donadi:2012nv}. A number of authors \cite{Uchiyama:1996va} --
\cite{Bertlmann:2001ea} try to adapt Wigner inequalities for 
oscillations of pseudoscalar mesons, usually the neutral $K$--mesons.
First, these adapted inequalities are studied in terms of
``flavour''--``$CP$-violation''--``states with defined masses and
lifetimes''. This idea was introduced in \cite{Uchiyama:1996va}. The
time-dependence is included by substitution of probabilities
calculated in the framework of quantum mechanics. In this case Wigner
inequalities become inequalities among the parameters $\varepsilon$
and $\varepsilon'$ of $CP$--violation. The violation of
these inequalities is small and is currently beyond experimental
reach \cite{Uchiyama:1996va}. Secondly, there are attempts to
include additional correlation functions which depend on 
time difference \cite{Bertlmann:2001ea}. A third way, introduced in
\cite{Privitera:1992xc} -- \cite{Foadi:2000zz}, is based on the
requirements of the causality principle and locality; however the
obtained inequalities are not general and are suitable only for the
specific situation of the oscillations of neutral mesons. Finally, in
\cite{Bertlmann:2001ea} -- \cite{Donadi:2012nv}, special versions
of time-dependent inequalities in the form \cite{Clauser:1969ny} are
introduced, but there are certain difficulties with their violation in
quantum mechanics.

To demonstrate the distinction between (\ref{W-B-4})
and (\ref{W-B-2}), we will apply them to the problem of oscillations
of neutral pseudoscalar mesons $M = \{K,\, D,\, B_q \}$, $q = \{ d,\,
s\}$. In this case the static inequalities (\ref{W-B-2}) are either not
violated at all, or the scale of the violation is beyond experimental
reach \cite{Uchiyama:1996va}. The violation of the time-dependent
inequality (\ref{W-B-4}), on the other hand, can be significantly
enchanced by a proper choice of parameters and hence allows
experimental tests.

\section{Static Wigner inequalities for oscillations of neutral
  pseudoscalar mesons}

\label{sec:1}

The key idea of the static Bell inequalities for the task at hand
was suggested in 
\cite{Privitera:1992xc,Uchiyama:1996va,Bramon:1998nz} and 
developed in \cite{Bertlmann:2001sk,Bertlmann:2001ea,Bramon:2005mg}. 

The essence of the idea is that there are three naturally provided
``directions'' whose projection operators do not commute.
The first is the flavour of a pseudoscalar meson. For example, for
the $B_q$--mesons, we consider the projections onto the states
$\ket{B_q} =\ket{\bar b q}$ and $\ket{\bar B_q} = \ket{b \bar q}$. We
 define the operators for charge ($\hat C$) and spatial ($\hat
P$) conjugation onto the states in the flavour space as 
\begin{eqnarray}
\hat C \hat P\,\ket{M}\, =\, e^{i\alpha} \ket{\bar
  M}\quad\textrm{and}\quad\hat C \hat P\,\ket{\bar M}\, =\, e^{-i
  \alpha} \ket{M}, \nonumber
\end{eqnarray}
where the $\alpha$ is a non-physical arbitrary real phase of 
$CP$--violation. This phase should be excluded from any experimentally
testable inequalities.

The second ``direction'' is the states with defined values of 
$CP$--parity, i.e. the states
\begin{eqnarray}
\ket{M_1} = \frac{1}{\sqrt{2}}\,\left ( \ket{M} + e^{i\alpha} \ket{\bar M}\right ), \quad 
\ket{M_2} = \frac{1}{\sqrt{2}}\,\left ( \ket{M} - e^{i\alpha} \ket{\bar M}\right ), \nonumber
\end{eqnarray}
which have positive and negative $CP$--parity accordingly. 

The third ``direction'' is defined by the states with fixed
values of mass and lifetime
\begin{eqnarray}
\ket{M_L} = p\left ( \ket{M} + e^{i \alpha}\,\frac{q}{p}\, \ket{\bar M}\right ) \quad\textrm{and}\quad
\ket{M_H} = p\left (\ket{M} - e^{i \alpha}\,\frac{q}{p}\, \ket{\bar M} \right ). \nonumber
\end{eqnarray}
The latter two states are the proper vectors of the non-hermitian hamiltonian
(for which $CPT$--symmetry is preserved):
$$
\hat H =  
\left (
\begin{array}{lr}
    \mathcal{H}              &  H_{12}\, e^{-i \alpha}\\
    H_{21}\, e^{i \alpha} &   \mathcal{H}
\end{array}
\right )\, =\,
\left (
\begin{array}{lr}
    m - \nicefrac{i}{2}\,\Gamma                                                                   &   \left ( m_{12} - \nicefrac{i}{2}\,\Gamma_{12} \right )\, e^{-i \alpha}\\
    \left ( m_{12}^* - \nicefrac{i}{2}\,\Gamma_{12}^* \right )\, e^{i \alpha} &   m - \nicefrac{i}{2}\,\Gamma
\end{array}
\right ),
$$
with the proper values 
\begin{eqnarray}
&& E_L  = m_L - \nicefrac{i}{2}\, \Gamma_L =   \mathcal{H} - \sqrt{H_{12} H_{21}} =   \mathcal{H} + \nicefrac{q}{p}\, H_{12}\quad \textrm{and} \nonumber \\
&& E_H = m_H - \nicefrac{i}{2}\, \Gamma_H  =  \mathcal{H} + \sqrt{H_{12} H_{21}} =   \mathcal{H} - \nicefrac{q}{p}\, H_{12} \nonumber
\end{eqnarray}
accordingly (here and subsequently we use the natural system of units
in which  $\hbar = c
=1$).  The states $\ket{M_L}$ and $\ket{M_H}$ are not orthogonal to
each other. The complex coefficients $p$ and $q$ are subjected to the
standard normalization condition:
\begin{eqnarray}
\label{pq-normirovka}
\bracket{M_L}{M_L}=\bracket{M_H}{M_H}=|p|^2 + |q|^2 = 1. 
\end{eqnarray}
We define
\begin{eqnarray}
&&  \Delta M = M_H - M_L = -\, 2\, \textrm{Re}\,\left ( \frac{q}{p}\, H_{12}\right ), \nonumber \\
&& \Delta \Gamma = \Gamma_H - \Gamma_L = 4\, \textrm{Im}\, \left ( \frac{q}{p}\, H_{12}\right ). \nonumber
\end{eqnarray}
Note that the definition of $\Delta \Gamma$ in the current work is
oppositely signed relative to the definition in 
\cite{Beringer:1900zz}. 

To automatically satisfy (\ref{pq-normirovka}), we introduce a new
variable $\beta$, for which 
\begin{eqnarray}
| p | = \cos\beta ;\qquad |q|=\sin\beta ;\qquad \textrm{and}\qquad
  \frac{q}{p} = \tg \beta \, e^{i \zeta} \equiv r  e^{i \zeta}, \qquad
  \beta \in \left [ 0,\, \pi/2 \right ]. \nonumber
\end{eqnarray}
Then
\begin{eqnarray}
\ket{M_L} = p \left ( \ket{M} + e^{i ( \alpha + \zeta )}\tg \beta \ket{\bar M} \right )   \quad\textrm{and}\quad  
\ket{M_H} = p \left ( \ket{M} - e^{i ( \alpha + \zeta )}\tg \beta \ket{\bar M} \right ). \nonumber
\end{eqnarray}

Decay of a neutral vector state $1^{-\, -}$  into an $M \bar
M$--pair (e.g. $\phi (1020) \to K \bar K$ or $\Upsilon (4S) \to
B \bar B$) defines a flavour-entangled wave function of 
the $M \bar M$--system at $t = t_0$:
\begin{eqnarray}
\label{correlationBbarB-t=0}
\ket{\Psi (t_0)} = \frac{1}{\sqrt{2}}
\left (\ket{M}^{(2)} \ket{\bar M}^{(1)}\, -\,\ket{\bar M}^{(2)} \ket{M}^{(1)}\right ).
\end{eqnarray}
The distinction between the first and the second meson can be provided by
their direction in the experimental device, see for example 
\cite{Uchiyama:1996va,Bertlmann:2001sk,Bertlmann:2001ea}. 

We obtain the static Wigner inequalities following the logic of
\cite{Bertlmann:2001ea}. We make the correspondence
$a_+ \to  M_1$, $a_- \to M_2$,
$b_+ \to \bar M$, $b_- \to M$, $c_+ \to M_H$ and $c_- \to M_L$. Then 
(\ref{W-B-2}) becomes:
\begin{eqnarray}
\label{W-B-2-BbarB-1}
w(M_1^{(2)},\, \bar M^{(1)},\, t_0)\,\le\, w(M_1^{(2)},\, M_H^{(1)},\, t_0) + w(M_H^{(2)},\, \bar M^{(1)},\, t_0).
\end{eqnarray} 
Substitution of probabilities (\ref{w-BbarB-I})  from Appendix
\ref{sec:A} into (\ref{W-B-2-BbarB-1}) leads to:
\begin{eqnarray}
\label{W-B-2-pq-1}
|q|^2 - |p|^2 \le \left | p + q\right |^2.
\end{eqnarray}
As there is no unambiguous correspondence between the projections of
the meson states onto various ``directions'' and the projections from
(\ref{W-B-2}), we set $b_+ \to M$ and $b_- \to \bar M$ while
keeping the $a_{\pm}$ and  $c_{\pm}$. Then (\ref{W-B-2}) becomes:
\begin{eqnarray}
\label{W-B-2-BbarB-2}
w(M_1^{(2)},\, M^{(1)},\, t_0)\,\le\, w(M_1^{(2)},\, M_H^{(1)},\, t_0) + w(M_H^{(2)},\, M^{(1)},\, t_0),
\end{eqnarray} 
and, taking into account  (\ref{w-BbarB-I}):
\begin{eqnarray}
\label{W-B-2-pq-2}
|p|^2 - |q|^2 \le \left | p + q\right |^2.
\end{eqnarray}
One can merge (\ref{W-B-2-pq-1}) and (\ref{W-B-2-pq-2}) as:
\begin{eqnarray}
\label{W-B-2-pq-3}
\left | |p|^2 - |q|^2 \right |\,\le\, \left | p + q\right |^2.
\end{eqnarray}
Now let $a_+ \to
M_2$ and $a_- \to M_1$, keeping the $b_{\pm}$ and $c_{\pm}$. Then from
(\ref{W-B-2}) follows 
\begin{eqnarray}
\label{W-B-2-BbarB-3}
w(M_2^{(2)},\, \bar M^{(1)},\, t_0)\,\le\, w(M_2^{(2)},\, M_H^{(1)},\, t_0) + w(M_H^{(2)},\, \bar M^{(1)},\, t_0),
\end{eqnarray}
and, taking into account (\ref{w-BbarB-I}):
\begin{eqnarray}
\label{W-B-2-pq-4}
|q|^2 - |p|^2 \le \left | p - q\right |^2.
\end{eqnarray}
Finally let $a_+ \to  M_2$, $a_- \to M_1$, $b_+ \to M$, $b_- \to \bar
M$,  and $c_+ \to M_H$, $c_- \to M_L$. Then, from (\ref{W-B-2}) it
follows that 
\begin{eqnarray}
\label{W-B-2-BbarB-4}
w(M_2^{(2)},\, M^{(1)},\, t_0)\,\le\, w(M_2^{(2)},\, M_H^{(1)},\, t_0) + w(M_H^{(2)},\, M^{(1)},\, t_0),
\end{eqnarray} 
and, in turn, 
\begin{eqnarray}
\label{W-B-2-pq-5}
|p|^2 - |q|^2 \le \left | p - q\right |^2.
\end{eqnarray}
The inqualities (\ref{W-B-2-pq-4}) and  (\ref{W-B-2-pq-5}) can be merged into
\begin{eqnarray}
\label{W-B-2-pq-6}
\left | |p|^2 - |q|^2 \right |\,\le\, \left | p - q\right |^2.
\end{eqnarray}
The inequalities (\ref{W-B-2-pq-3}) -- (\ref{W-B-2-pq-6}) represent
the 
full set of the static Wigner inequalities for the oscillations of
neutral mesons. The set is obtained from all possible correspondences
between the $a_{\pm}$, $b_{\pm}$, and $c_{\pm}$ and the projection of
the meson states onto the ``directions'' of flavour, $CP$, and the
states with fixed masses and lifetimes.
It is obvious that 
(\ref{W-B-2-pq-3}) and (\ref{W-B-2-pq-6}) can not be violated
simultaneously -- one can sum  
(\ref{W-B-2-pq-3}) and (\ref{W-B-2-pq-6}).
However there are no physical arguments to prefer 
(\ref{W-B-2-pq-3}) over (\ref{W-B-2-pq-6}) or vice versa.

Using (\ref{w-BbarB-I}) it can be shown that 
(\ref{W-B-2-pq-3}) and (\ref{W-B-2-pq-6}) reduce to
\begin{eqnarray}
\label{W-B-2-pq-sinus-3}
\left | \cos 2 \beta \right |\, \pm\, \cos\zeta\, \sin ( 2 \beta )\,\le\,1.
\end{eqnarray}
Inequality (\ref{W-B-2-pq-sinus-3}) does not contain the unphysical phase
$\alpha$, as must be the case for any experimentally testable relation
between the observables in quantum theory. 

We now check whether  (\ref{W-B-2-pq-3}) is violated in
systems of neutral pseudoscalar mesons. In Table 
I the current experimental values of $\left|\nicefrac{q}{p}\right
|^{exp}_M$ are shown. For all the neutral mesons 
$\left|\nicefrac{q}{p}\right |^{exp}_M \approx 1$, i.e. $\beta
\approx \beta_0 = \nicefrac{\pi}{4}$. For the $D$--mesons, $\left|\nicefrac{q}{p}\right
|^{exp}_D$ is not well measured, however within the experimental
uncertainties it is consistent with one. 

For $\beta = \beta_0 = \nicefrac{\pi}{4}$ (i.e. without oscillation-induced
$CP$--violation), (\ref{W-B-2-pq-sinus-3}) reduces to the
trivial inequality
\begin{eqnarray}
\label{W-B-2-pq-sinus-4}
 \left | \cos\zeta \right | \,\le\, 1,
\end{eqnarray}
which is not violated for any value of the phase $\zeta$.

\begin{table}
\label{table:parameters}
\caption{Experimental values of the oscillations and the
  $CP$--violation for the neutral pseudoscalar mesons
  \protect\cite{Beringer:1900zz}.
The minus sign of $\Delta\Gamma$ is due to the difference of the
definitions between the current work and
\protect\cite{Beringer:1900zz}. The dimensionless variable 
$\lambda = \nicefrac{\Delta M}{\Delta\Gamma}$.}
\centering
\begin{tabular}{||c|c|r|c|c||}
\hline
\hline
Meson & $\Delta\Gamma~ (\textrm{MeV})$ & $\Delta M ~(\textrm{MeV})$ & $\tan\beta \equiv \left|\nicefrac{q}{p}\right |^{exp}_M$ & $\lambda$\\ 
\hline\hline 
$B^{0}_{s}$ & $-\, 6.0\times 10^{-11}$ & $1.2\times 10^{-8}$& $1.0039\pm 0.0021$                  & $- 0.2 \times 10^{3}$ \\ 
$K^{0}$        &  $-\, 7.3\times 10^{-12}$ & $3.5\times 10^{-12}$ & $0.99668\pm 0.00004$          & $- 4.8 \times 10^{-1}$ \\ 
$D^{0}$        & $ -\, 2.1\times 10^{-11}$ & $ -\, 6.3\times 10^{-12}$ & $0.92^{+0.12}_{-0.09}$  & $0.3$ \\
\hline
\hline
\end{tabular} 
\end{table}

However (\ref{W-B-2-pq-sinus-4}) does not mean that the static
inequalities (\ref{W-B-2-pq-3}) and (\ref{W-B-2-pq-6})
are never violated. Due to the $CP$--violation, the angle $\beta$ is
slightly different from $\beta_0 =\nicefrac{\pi}{4}$. 

For $K$--mesons violation of (\ref{W-B-2-pq-3}) was
demonstrated in \cite{Uchiyama:1996va}. In the case at hand, 
the coefficients $p$ and $q$ are defined through the $CP$--violation
parameter $\varepsilon$ as:
\begin{eqnarray}
p = \frac{1}{\sqrt{2}}\,\frac{1 + \varepsilon}{\sqrt{1 + |\varepsilon|^2}}\quad\textrm{and}\quad
q = \frac{1}{\sqrt{2}}\,\frac{1 - \varepsilon}{\sqrt{1 + |\varepsilon|^2}}. \nonumber
\end{eqnarray}
Then (\ref{W-B-2-pq-3}) becomes 
\begin{eqnarray}
\label{W-B-2-pq-varepsilon-2}
\left | \textrm{Re}\left ( \varepsilon \right ) \right | \,\le\, 1.
\end{eqnarray}
and (\ref{W-B-2-pq-6}) becomes
\begin{eqnarray}
\label{W-B-2-pq-varepsilon-1}
\left | \textrm{Re}\left ( \varepsilon \right ) \right | \,\le\, |\varepsilon|^2,
\end{eqnarray}
which corresponds to (16) from \cite{Uchiyama:1996va} if one neglects
the corrections $\sim |\varepsilon|^2$ and the modulus on the left
hand side. 
The inequality (\ref{W-B-2-pq-varepsilon-2}) is never violated,
as $|\varepsilon| \sim 10^{-3}$ \cite{Beringer:1900zz}, leading to
the upper limit $|\textrm{Re}\left ( \varepsilon \right )| \sim
10^{-3}$. Inequality (\ref{W-B-2-pq-varepsilon-1}) should
be strongly violated, as $\varphi_{\varepsilon} = (43.52 \pm 0.05)^o$
\cite{Beringer:1900zz}. However due to the smallness of 
$CP$--violation in neutral $K$--mesons, direct experimental tests
of (\ref{W-B-2-BbarB-3}) and (\ref{W-B-2-BbarB-4}) are not possible 
\cite{Uchiyama:1996va}.

\section{Time-dependent Wigner inequalities for the oscillations of
  neutral pseudoscalar mesons}
\label{sec:2}

We now consider time-dependent Wigner inequalities (\ref{W-B-4})
for neutral pseudoscalar meson systems. Note that the 
normalization of probability to unity was not used in the derivation of (\ref{W-B-4}).
Hence (\ref{W-B-4}) is valid for a unstable particles, whose
state vector normalization is time-dependent. 

The time evolution of the states $\ket{M_L}$ è $\ket{M_H}$ is given by:
\begin{eqnarray}
\ket{M_L (t)} &=&  e^{- i E_L \,\Delta t}  \ket{M_L} = e^{- i m_L \,\Delta t - \Gamma_L\, \Delta t /2} \ket{M_L}, \\
\ket{M_H (t)} &=&  e^{- i E_H \,\Delta t}  \ket{M_H} = e^{- i m_H \,\Delta t - \Gamma_H\, \Delta t / 2} \ket{M_H}, \nonumber
\end{eqnarray}
where $\Delta t = t - t_0$.
This leads to the evolution of the states $\ket{M (t)}$ è $\ket{\bar M (t)}$
as:
\begin{eqnarray}
\left \{
\begin{array}{l}
\displaystyle \ket{M(t)} = g_+(\Delta t) \ket{M}\, -\, e^{i \alpha}\,\frac{q}{p}\, g_-(\Delta t) \ket{\bar M} \\
\displaystyle \ket{\bar M (t)} = g_+(\Delta t) \ket{\bar M} - e^{-i \alpha}\,\frac{p}{q}\, g_-(\Delta t) \ket{M}
\end{array}
\right . . \nonumber
\end{eqnarray}
We then obtain the evolution of the states $\ket{M_1(t)}$ and
$\ket{M_2(t)}$ as:
\begin{eqnarray}
\ket{M_1(t)} = \frac{1}{\sqrt{2}}
\left (
\left ( 
g_+(\Delta t) - \,\frac{p}{q}\, g_-(\Delta t)
\right ) \ket{M}\, +\,
e^{i \alpha}\,\left ( 
g_+(\Delta t)\,  -\,\frac{q}{p}\, g_-(\Delta t)
\right ) \ket{\bar M}
\right ),
\nonumber \\
\ket{M_2(t)} = \frac{1}{\sqrt{2}}
\left (
\left ( 
g_+(\Delta t) + \,\frac{p}{q}\, g_-(\Delta t)
\right ) \ket{M}\, -\,
e^{i \alpha}\,\left ( 
g_+(\Delta t)\,  +\,\frac{q}{p}\, g_-(\Delta t)
\right ) \ket{\bar M}
\right ),
\nonumber
\end{eqnarray}
where $\displaystyle g_{\pm}(\tau) = \frac{1}{2}\,\left ( e^{-i E_H
    \tau} \pm e^{- i E_L \tau}\right )$. 
For the functions $g_{\pm}(\tau)$,
the following is satisfied: 
\begin{eqnarray}
&& \left | g_{\pm}(\tau) \right |^2 = \frac{e^{-\Gamma \tau}}{2}\,
\left (
\ch \left ( \frac{\Delta\Gamma\, \tau}{2}\right ) \pm \cos \left ( \Delta M\, \tau\right ) \right ), \nonumber \\
&& g_+^* (\tau) g_-(\tau) = -\,\frac{e^{-\Gamma \tau}}{2}\,
\left (
\sh \left ( \frac{\Delta\Gamma\, \tau}{2}\right ) +  i\sin \left ( \Delta M\, \tau\right ) 
\right ),  \nonumber 
\end{eqnarray}
where $\Gamma = (\Gamma_H + \Gamma_L)/2$. Taking into account 
initial condition 
(\ref{correlationBbarB-t=0}), it is possible to write the wave
function of the $M\bar M$--pair at arbitrary time $t$:
\begin{eqnarray}
\label{correlationBbarB-t}
\ket{\Psi (t)} =  e^{-i \left ( m_H + m_L\right )\, \Delta t}\, e^{- \Gamma\, \Delta t}\,\ket{\Psi (t_0)}.
\end{eqnarray}
To simplify the subsequent calculations we from now on set $t_0
=0$, so $\Delta t \equiv t$. 

To demonstrate the advantage of (\ref{W-B-4}) over
(\ref{W-B-2}), we first consider the case of  $\beta = \beta_0 =
\nicefrac{\pi}{4}$, when there is no oscillation-induced
$CP$--violation. The static inequality 
(\ref{W-B-2}) becomes the never-violated inequality
 (\ref{W-B-2-pq-sinus-4}). One can obtain a significant simplification
by considering the time-dependent (\ref{W-B-4})  with the additional
condition $\cos\zeta = \pm 1$. If one neglects 
$CP$--violation, then for $K$--mesons, 
$
\displaystyle \left (\frac{q}{p} \right )_K = \frac{1 - \epsilon}{1 + \epsilon} \approx 1
$,
so $\cos\zeta_K = 1$. For $B_q$--mesons the effective
Hamiltonian of the oscillations is proportional to $\left ( V_{tb}
  V_{tq}^*\right)^2$ \cite{Buras:2001ra}. Then:
$$
\displaystyle \left (\frac{q}{p} \right )_{B_q} = -\,\frac{H_{21}}{\sqrt{H_{12}\, H_{21}}} \approx \, -\,
\left (\frac{V_{tb}^* V_{tq}}{\left | V_{tb}^* V_{tq} \right |} \right )^2 = -1,
$$ 
hence $\cos\zeta_{B_q} = -1$. For $D$--mesons 
the experimental results of  BaBar \cite{delAmoSanchez:2010xz} and Belle
\cite{Abe:2007rd} are in accordance with the assumption $\cos\zeta_D =
1$, in which case the condition $\cos\zeta = \pm 1$ is well justified.

Table II shows 
time-dependent Wigner inequalities for all possible correspondences
between the dichotomic variables $a_{\pm}$, $b_{\pm}$, $c_{\pm}$ and
the projections of meson states onto the ``directions'' of 
flavour, $CP$, and states with fixed masses and lifetimes.

All calculations are performed using formulas (\ref{w-BbarB-I}) and
(\ref{w-BbarB-II}) with the approximation $\beta = \beta_0 =
\nicefrac{\pi}{4}$ and $\cos\zeta = \pm 1$. The
experimental values of the oscillation parameters shown in Table I, 
and numerical estimates of the $\cos \zeta_M$ suggest the optimal
choice of sets N5 and N6 from Table II for studying the violation of
 (\ref{W-B-4}) in $K$- and $D$--mesons. For studying the
violation of (\ref{W-B-4}) in oscillations of neutral
$B_{d,s}$--mesons one should  choose sets N7 and N8.

\begin{table}
\centering
\caption{ Time-dependent Wigner inequalities  (\protect\ref{W-B-4})
  for neutral pseudoscalar mesons with the approximation $\beta =
  \beta_0 =  \nicefrac{\pi}{4}$ and $\cos\zeta = \pm 1$. All possible
  correspondences between the dichotomic variables $a_{\pm}$, $b_{\pm}$,
  $c_{\pm}$ and the projections of the meson states onto the ``directions''
  of flavour, $CP$, and the states with fixed masses and
  lifetimes are shown.
\hfill\label{table:TDBU}}
\bigskip
{\setlength{\extrarowheight}{4pt}
\begin{tabular}{||c|c|c|c||}
\hline
\hline
N & Correspondence of  & Time-dependent                  & Violation \\ 
    & the variables             & Wigner inequalities             & conditions         \\
\hline\hline 
1 & $a_+ \to M_1$,
      $b_+ \to \bar M$, 
      $c_+ \to M_H$,    & $1 \le e^{- \Delta \Gamma\, t}$ & if $\Delta \Gamma \ge 0$ \\
  & $a_- \to M_2$,
      $b_- \to M$,
      $c_- \to M_L$       &  when $\cos\zeta = -1$                 &                                                \\
\hline
2 & $a_+ \to M_1$,
       $b_+ \to M$, 
       $c_+ \to M_H$,    & $1 \le e^{- \Delta \Gamma\, t}$ & if $\Delta \Gamma \ge 0$ \\
  &  $a_- \to M_2$,
       $b_- \to \bar M$,
       $c_- \to M_L$       &  when $\cos\zeta = -1$                 &                                                \\
\hline
3 & $a_+ \to M_2$,
       $b_+ \to \bar M$, 
       $c_+ \to M_H$,    & $1 \le e^{- \Delta \Gamma\, t}$ & if $\Delta \Gamma \ge 0$ \\
 &   $a_- \to M_1$,
       $b_- \to M$,
       $c_- \to M_L$       &  when $\cos\zeta = +1$                 &                                                \\
\hline
4 & $a_+ \to M_2$,
       $b_+ \to M$, 
       $c_+ \to M_H$,    & $1 \le e^{- \Delta \Gamma\, t}$ & if $\Delta \Gamma \ge 0$ \\
   & $a_- \to M_1$,
       $b_- \to \bar M$,
       $c_- \to M_L$       &  when $\cos\zeta = +1$                 &                                                \\
\hline
5 & $a_+ \to M_1$,
       $b_+ \to \bar M$, 
       $c_+ \to M_L$,    & $1 \le e^{\Delta \Gamma\, t}$ & if $\Delta \Gamma \le 0$ \\
 &   $a_- \to M_2$,
       $b_- \to M$,
       $c_- \to M_H$       &  when $\cos\zeta = +1$                 &                                                \\
\hline
6 & $a_+ \to M_1$,
       $b_+ \to M$, 
       $c_+ \to M_L$,    & $1 \le e^{\Delta \Gamma\, t}$ & if $\Delta \Gamma \le 0$ \\
 &   $a_- \to M_2$,
       $b_- \to \bar M$,
       $c_- \to M_H$       &  when $\cos\zeta = +1$                 &                                                \\
\hline
7 & $a_+ \to M_2$,
       $b_+ \to \bar M$, 
       $c_+ \to M_L$,    & $1 \le e^{\Delta \Gamma\, t}$ & if $\Delta \Gamma \le 0$ \\
 &   $a_- \to M_1$,
       $b_- \to M$,
       $c_- \to M_H$       &  when $\cos\zeta = -1$                 &                                                \\
\hline
8 & $a_+ \to M_2$,
       $b_+ \to M$, 
       $c_+ \to M_L$,    & $1 \le e^{\Delta \Gamma\, t}$ & if $\Delta \Gamma \le 0$ \\
 &   $a_- \to M_1$,
       $b_- \to \bar M$,
       $c_- \to M_H$       &  when $\cos\zeta = -1$                 &                                                \\
\hline
\hline
\end{tabular} 
}
\end{table}

\section{$CP$--violation effects influencing the violation of the
  time-dependent Wigner inequalities}
\label{sec:3}

We now take into account all the $CP$--violation effects, i.e. the
case when $\beta \ne \beta_0$, and $\cos\zeta_M \ne \pm
1$. Then for the various sets from Table
\ref{table:TDBU}, the substitution of  (\ref{w-BbarB-I}) and
(\ref{w-BbarB-II}) into the time-dependent Wigner inequalities
(\ref{W-B-4}) results in eight inequalities. They can be reduced to:
\begin{eqnarray}
\label{W-B-2-gammaT}
	1 \,\leq\,\mathrm{R_N}(x,\, r,\, \zeta,\, \lambda).
\end{eqnarray}
The functions $\mathrm{R_N}$ depend on the dimensionless variables $x
= \Delta\Gamma t$, $\lambda = \nicefrac{\Delta M}{\Delta\Gamma}$,
the absolute value $r$, and the phase $\zeta$ of the ratio
$\nicefrac{q}{p}$.
The experimental values of $\Delta \Gamma$ are less or equal
to  $0$ for $K$--,
$D$--, and $B_s$--mesons, thus we consider only the functions $\mathrm{R_5}$ --
$ñ\mathrm{R_8}$. Analytical expressions for these functions are given
in Appendix \ref{sec:B}. Numerical values of the parameters used
for the analysis of the inequalities (\ref{W-B-2-gammaT})
are given in Table I.  %

\begin{figure}[p]
\begin{center}
	\begin{tabular}{c}
	\mbox{\epsfig{file=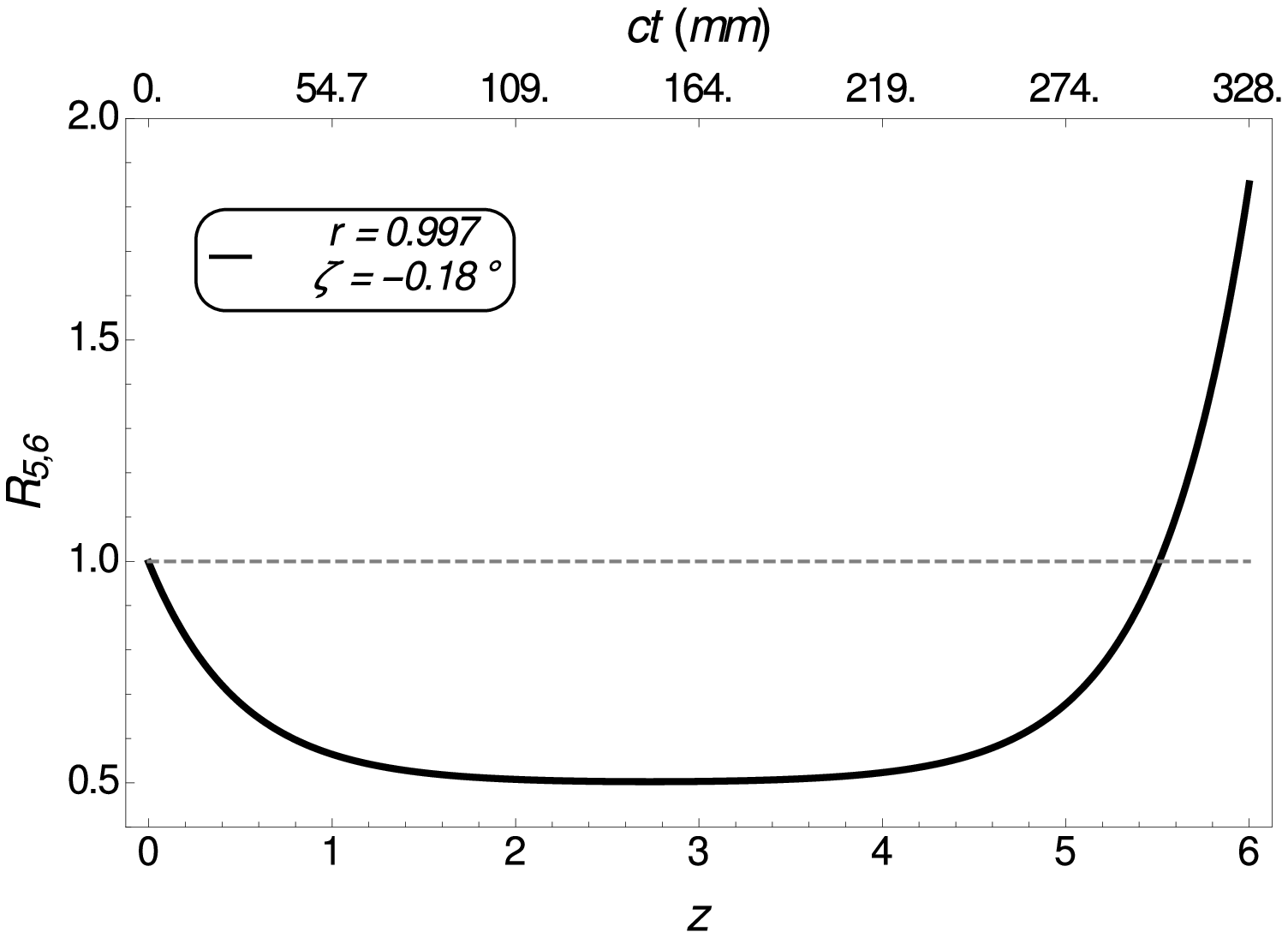,width=12.cm}}
	\end{tabular}
	\caption{\label{fig:RHS_K} Functions $\textrm{R}_{5,\, 6} (x,\,
      r,\,\zeta,\,\lambda)$ for neutral $K$--mesons (both functions
      are almost juxtaposed due to the high accuracy of the 
      $CP$--violation parameter $\varepsilon$). The scale  at the top corresponds
      to the variable $c \,t$ (mm); the bottom scale -- to the time in 
      units of the average lifetime $z = (\Gamma_H + \Gamma_L)\, t /2 =
      \Gamma \, t$, where $t$ is kaon rest frame time. 
}
\end{center}
\end{figure}

\begin{figure}[p]
\begin{center}
	\begin{tabular}{c}
	\mbox{\epsfig{file=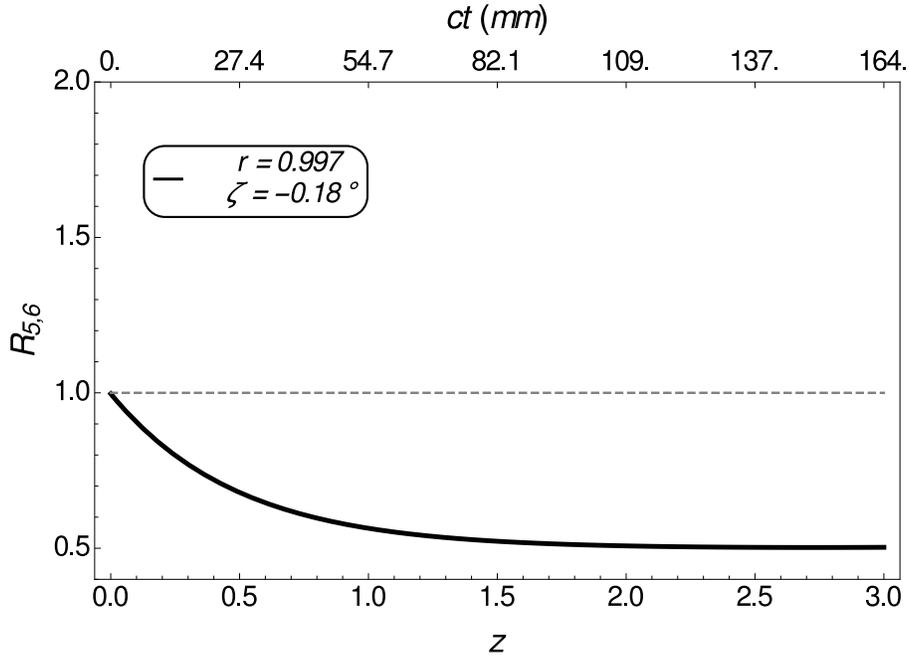,width=12.cm}}
	\end{tabular}
	\caption{\label{fig:RHS_K3} Functions $\textrm{R}_{5,\, 6} (x,\,
      r,\,\zeta,\,\lambda)$ for neutral $K$--mesons for $z
      \le 3$.} %
\end{center}
\end{figure}

We start with the system of neutral kaons. For $K$--mesons the
absolute values and phase of the $CP$ violation parameter $\varepsilon$ are
known with quite high precision. Hence $r$ and $\zeta$ are also well 
defined. 
In FIG. \ref{fig:RHS_K} the functions 
$\mathrm{R_5}(x,\, r,\, \zeta,\, \lambda)$ and $\mathrm{R_6}(x,\, r,\,
\zeta,\, \lambda)$ are shown. For kaons these functions are almost
identical for the experimentally allowed values of $r$ and $\zeta$ (e.g. with
$r = 0.997$  and $\zeta = - 0.18^o$, which are used in FIG
\ref{fig:RHS_K}).  The top scale corresponds to the variable
$c t$  (the decay length) in mm. The bottom scale corresponds to 
time measured in units of the average kaon lifetimes $\displaystyle
z = \frac{1}{2} \left ( \Gamma_H + \Gamma_L \right )\, t\,  = \,
\Gamma \, t$. Time $t$ is calculated in the $K$--meson rest frame.
In FIG. \ref{fig:RHS_K3} the functions 
$\mathrm{R_{5,\, 6}}(x,\, r,\, \zeta,\, \lambda) $ are shown for $z
\le 3$, i.e. in the most experimentally accessible area.

Time-dependent inequalities (\ref{W-B-2-gammaT}) are violated when 
$\mathrm{R_{5,\, 6}}(x,\, r,\, \zeta,\, \lambda) <
1$. FIG. \ref{fig:RHS_K}, shows that for the $K$--mesons this violation occurs
when $z \lesssim 5.5$, i.e., in the experimentally accessible area. 
For $z \gtrsim 5.5$ the inequalities (\ref{W-B-2-gammaT}) are not
violated. This interesting effect may be understood if one makes
an expansion of the functions 
$\mathrm{R_N}(x,\, r,\, \zeta,\, \lambda)$ by small parameters $\Delta
r = r-1$ and $\zeta$. As an example we obtain the expansion of the
function $\mathrm{R_5}(x,\, r,\, \zeta,\, \lambda)$  to second
order. We consider $\Delta r$ and $\zeta$ to be of the same order of
magnitude. Then:

\begin{eqnarray}
 \mathrm{R_5}(x,\, r,\, \zeta,\, \lambda)  & \approx& 
 \frac{1}{2}  \left(e^{x}+1\right) \, (1\, -\,\Delta r )\, + \nonumber  \\
 &+ &  \left ( 3\,\mathrm{ch}^2 \left(\frac{x}{2}\right) \,+\, \frac{3}{2}\,\sh \left(\frac{x}{2}\right)\, 
 	\ch \left( \frac{x}{2}\right ) \, -2\, \ch \left(\frac{x}{2}\right)\, \cos \left ( \lambda x \right ) \right ) \, (\Delta r)^2 \, +  \nonumber \\
  &+ & \left ( \frac{3}{2} \mathrm{ch} ^2\left(\frac{x}{2}\right)-\ch \left(\frac{x}{2}\right)\,\cos (\lambda x) \right  )\, \zeta^2\, -  
             \ch \left(\frac{x}{2}\right) \sin (\lambda x) \,\zeta\, \Delta r . \nonumber
\end{eqnarray}

At zeroth order in $\Delta r$ and $\zeta$, which corresponds to the
absence of $CP$--violation, the effect of restoration of the
inequality (\ref{W-B-2-gammaT}) at large values of $t$ (or $z$) does
not appear. That is, this effect is fully determined by the
$CP$--violation. A first order expansion is also not enough. The effect
appears when, at a particular value of
$z$ the second order contribution begins to be comparable to the
previous orders and the expansion is no longer valid.
Note that in the range of low $z$, which is the most
experimentally interesting, the approximation from Table 
II is thus proved to be acceptable. Similar properties can be observed in
the expansion of the function $\mathrm{R_6}(x,r,\zeta,\lambda)$.

\begin{figure}[p]
\begin{center}
	\begin{tabular}{ccc}
	\mbox{\epsfig{file=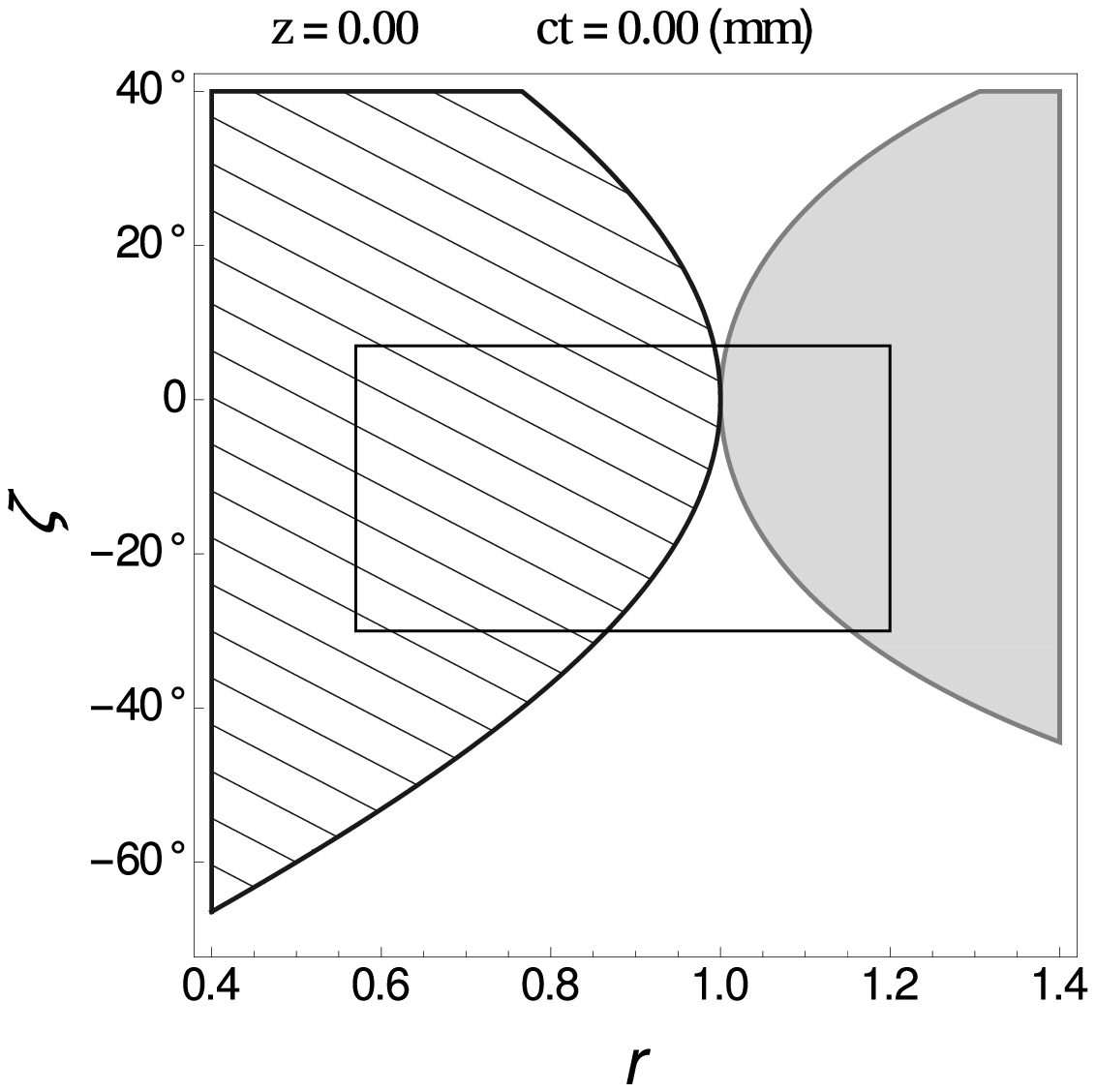,width=5.2cm}} & \mbox{\epsfig{file=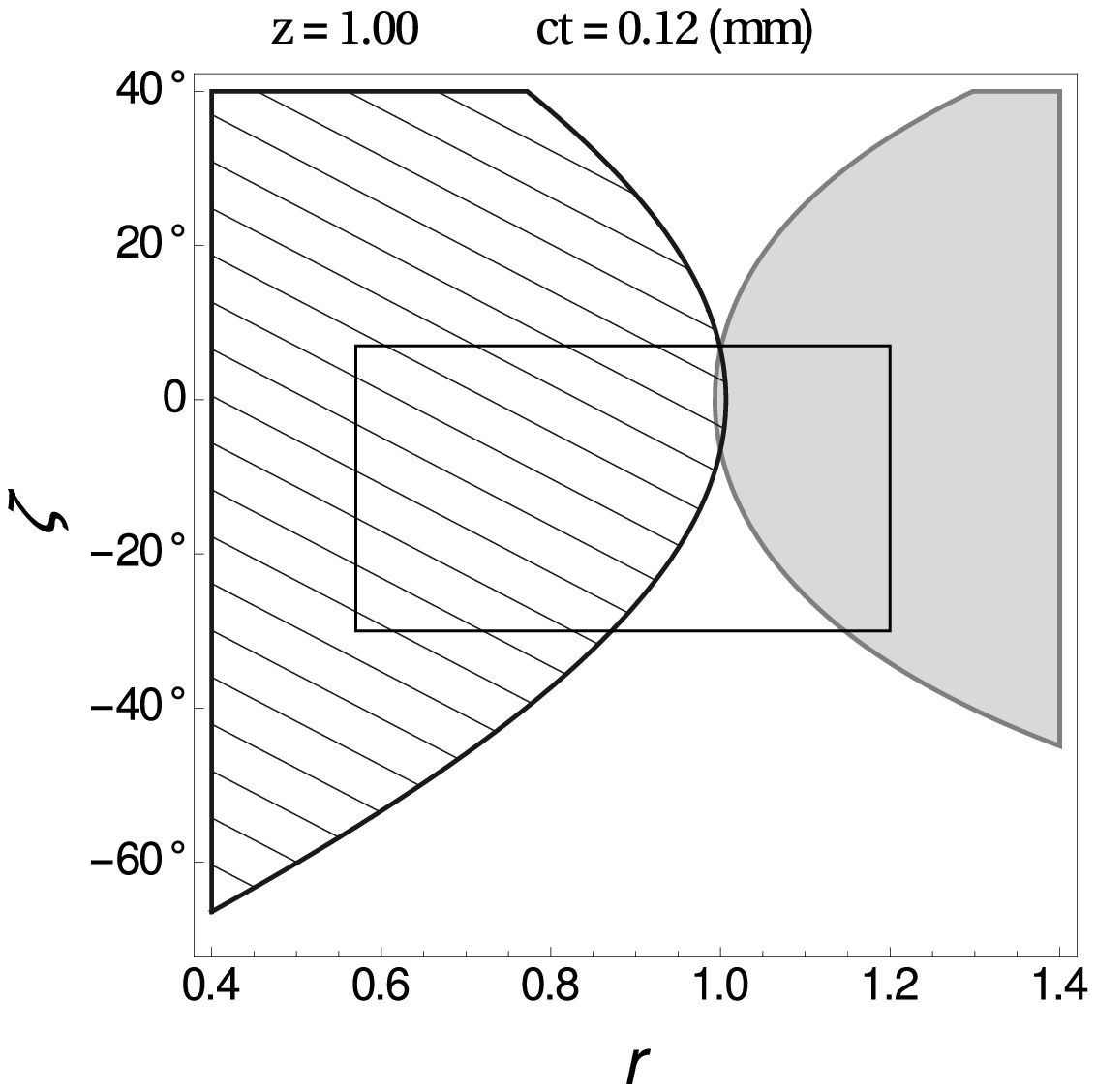,width=5.2cm}} & \mbox{\epsfig{file=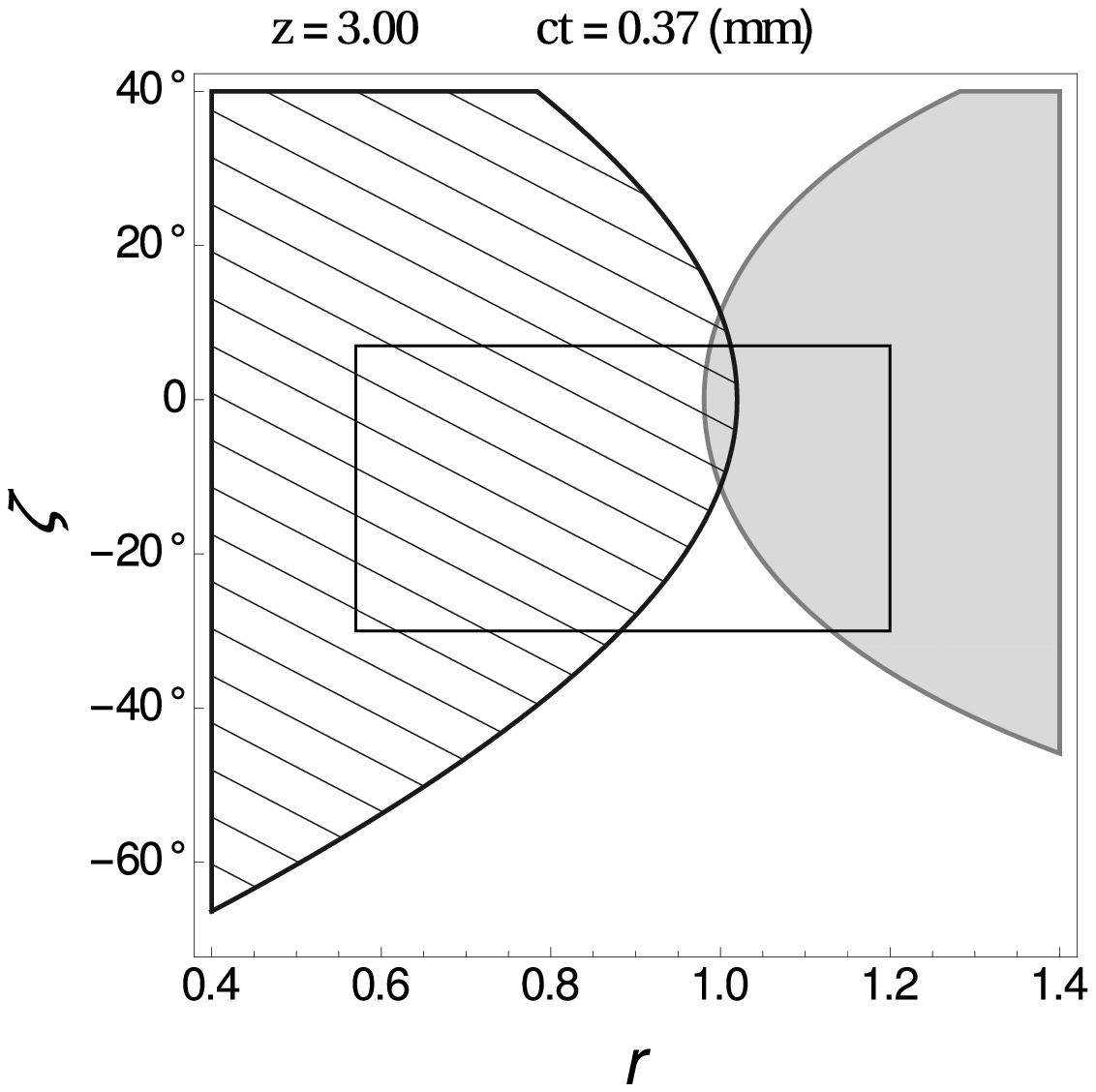,width=5.2cm}} \\
         \mbox{\epsfig{file=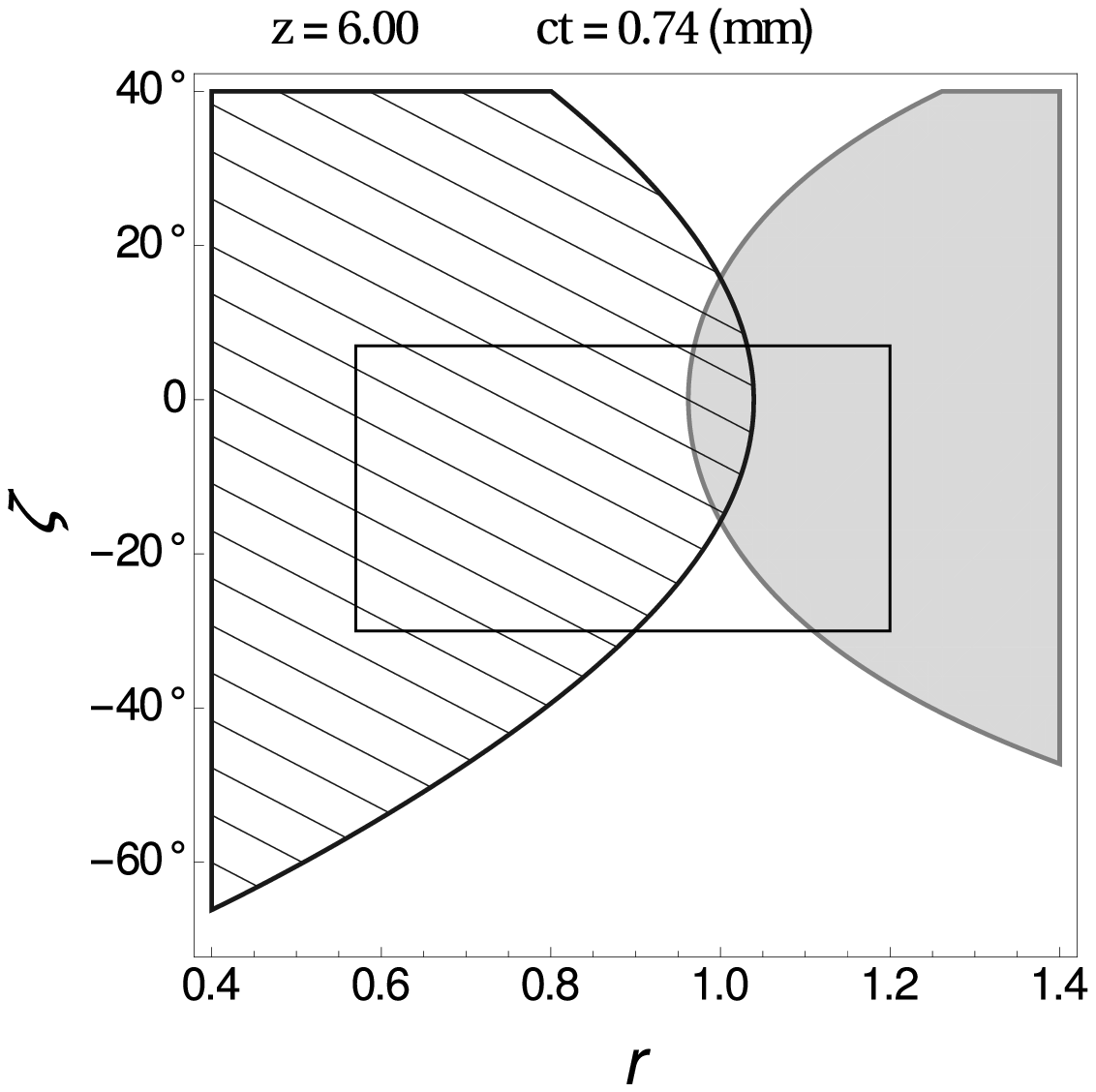,width=5.2cm}} & \mbox{\epsfig{file=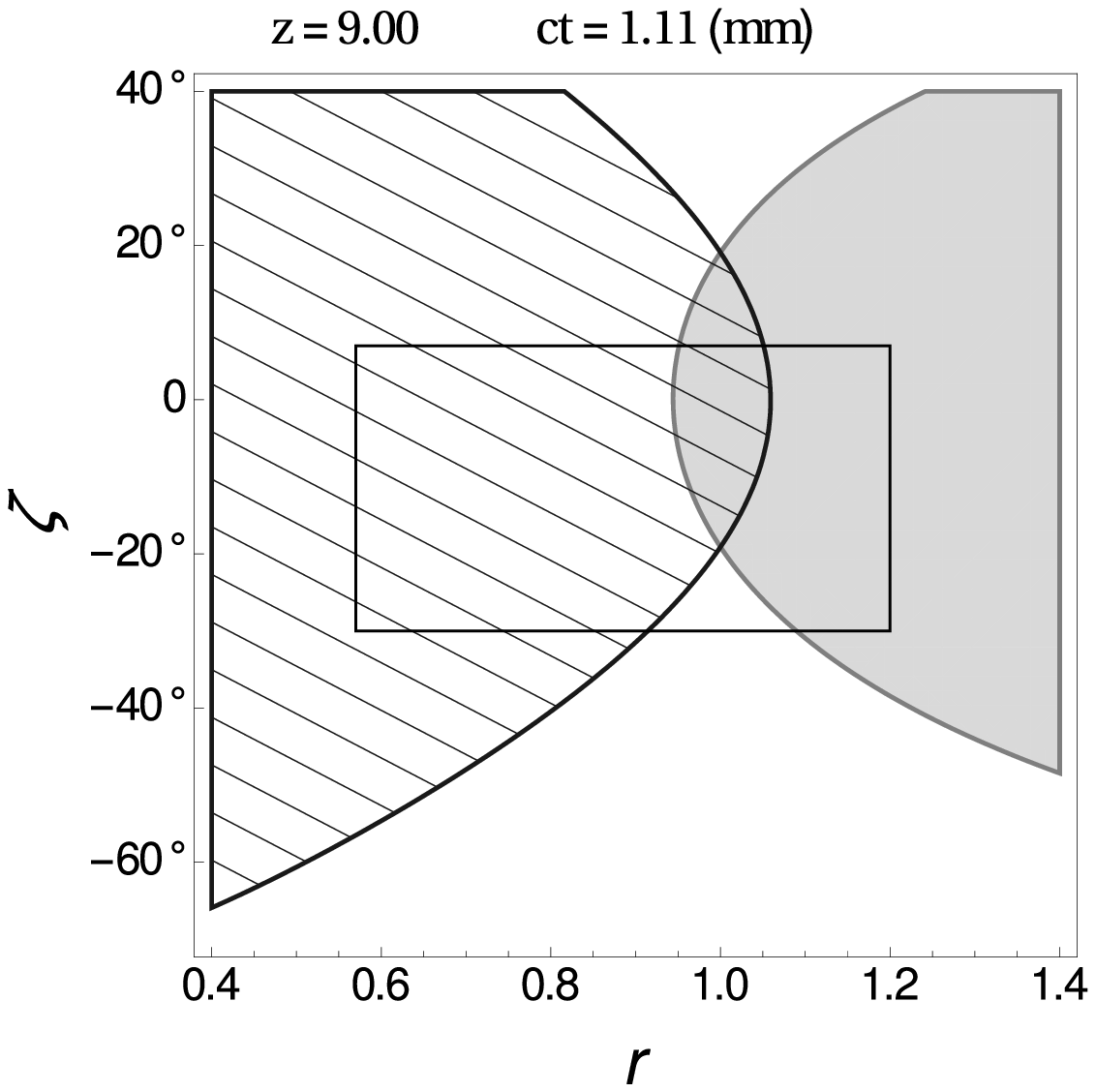,width=5.2cm}} & \mbox{\epsfig{file=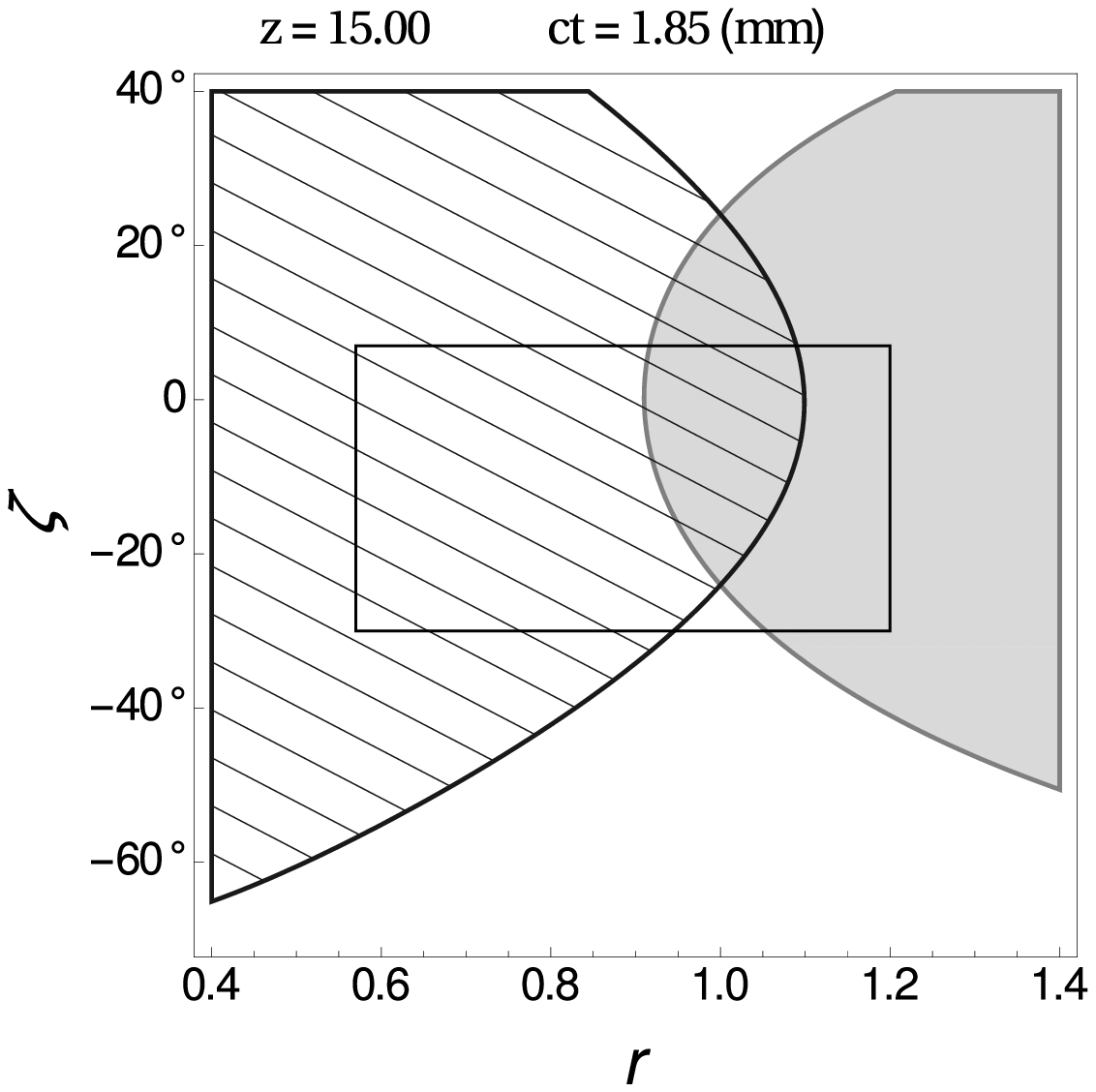,width=5.2cm}} \\
         \mbox{\epsfig{file=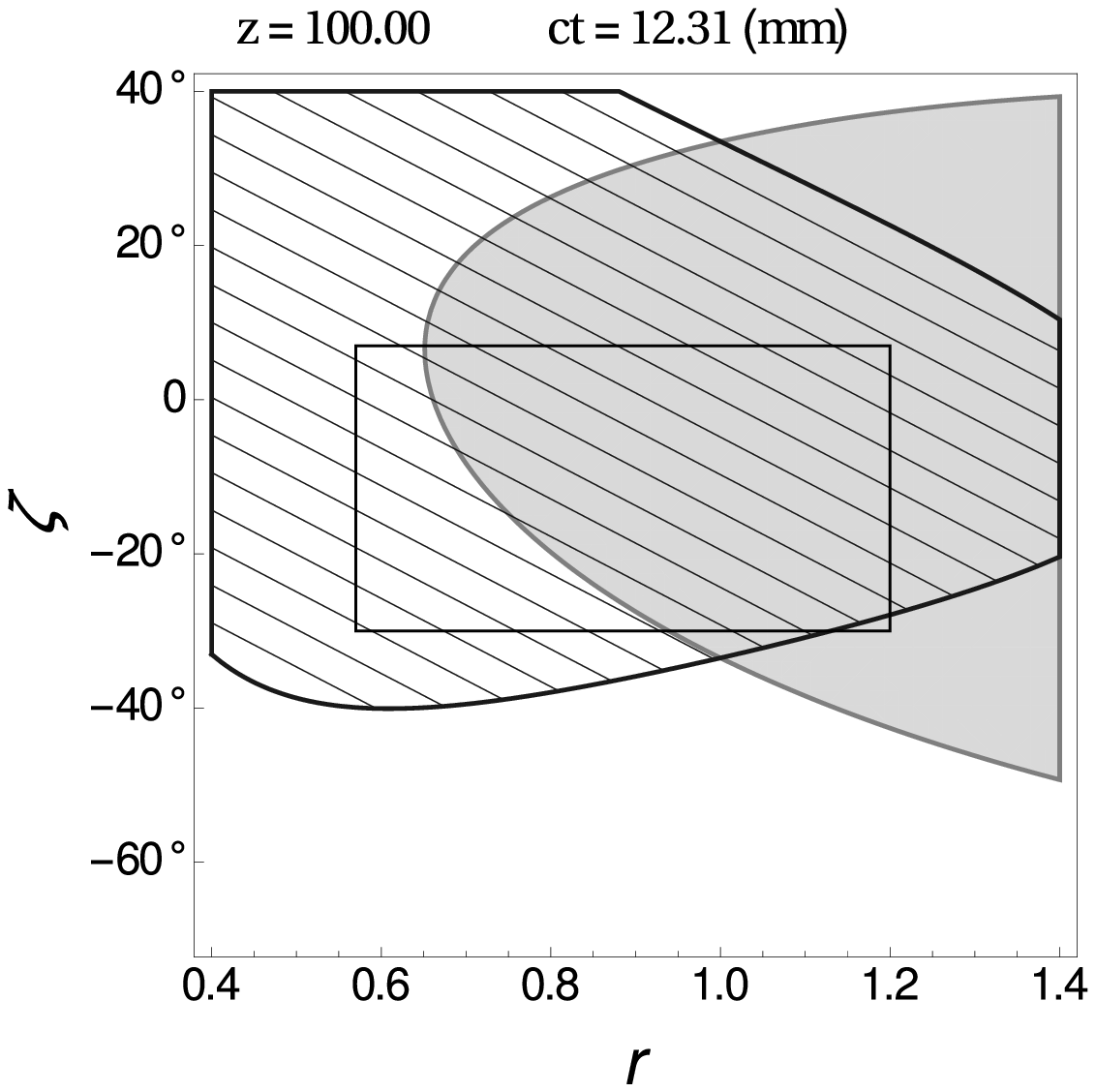,width=5.2cm}} & \mbox{\epsfig{file=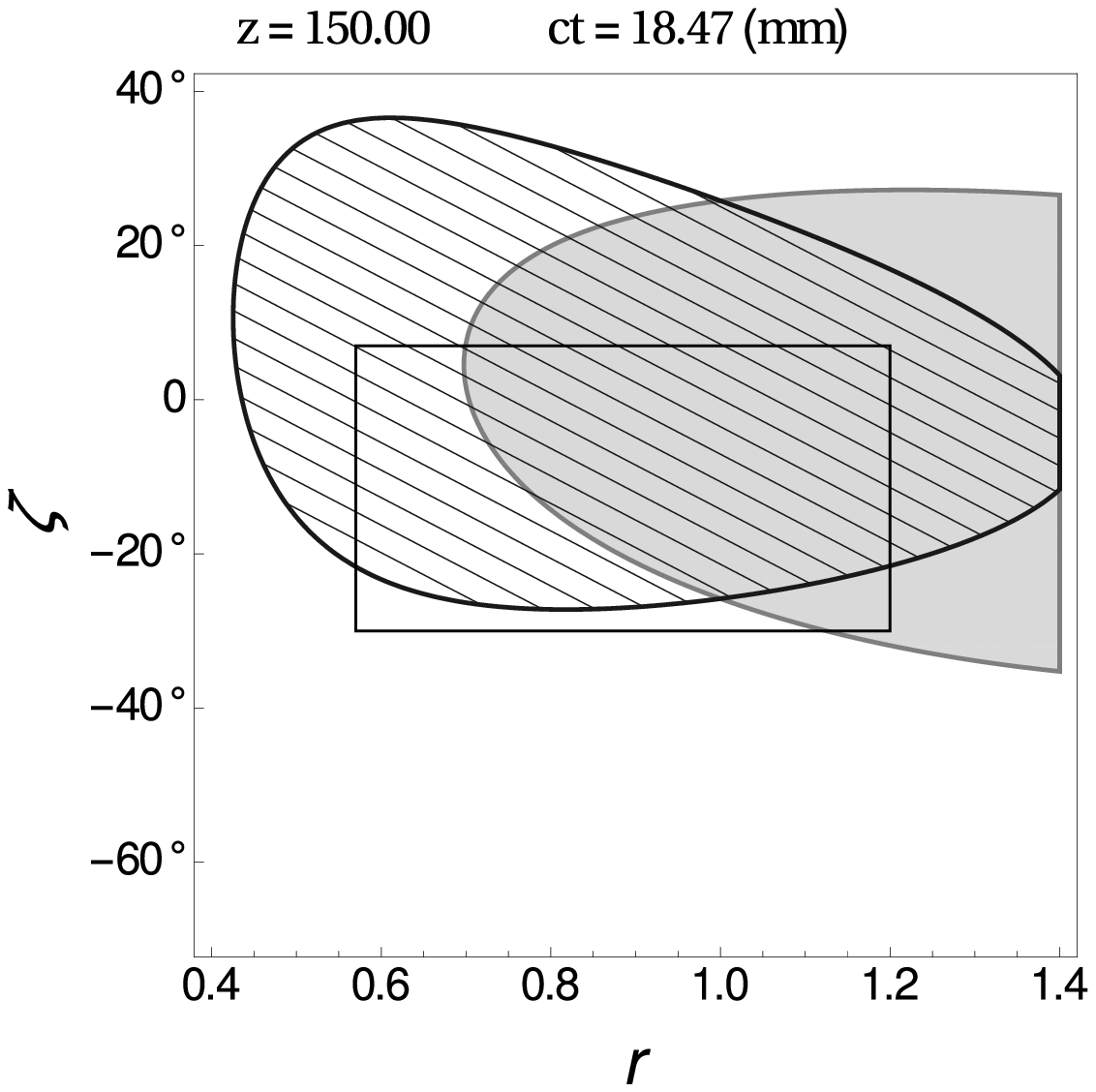,width=5.2cm}} & \mbox{\epsfig{file=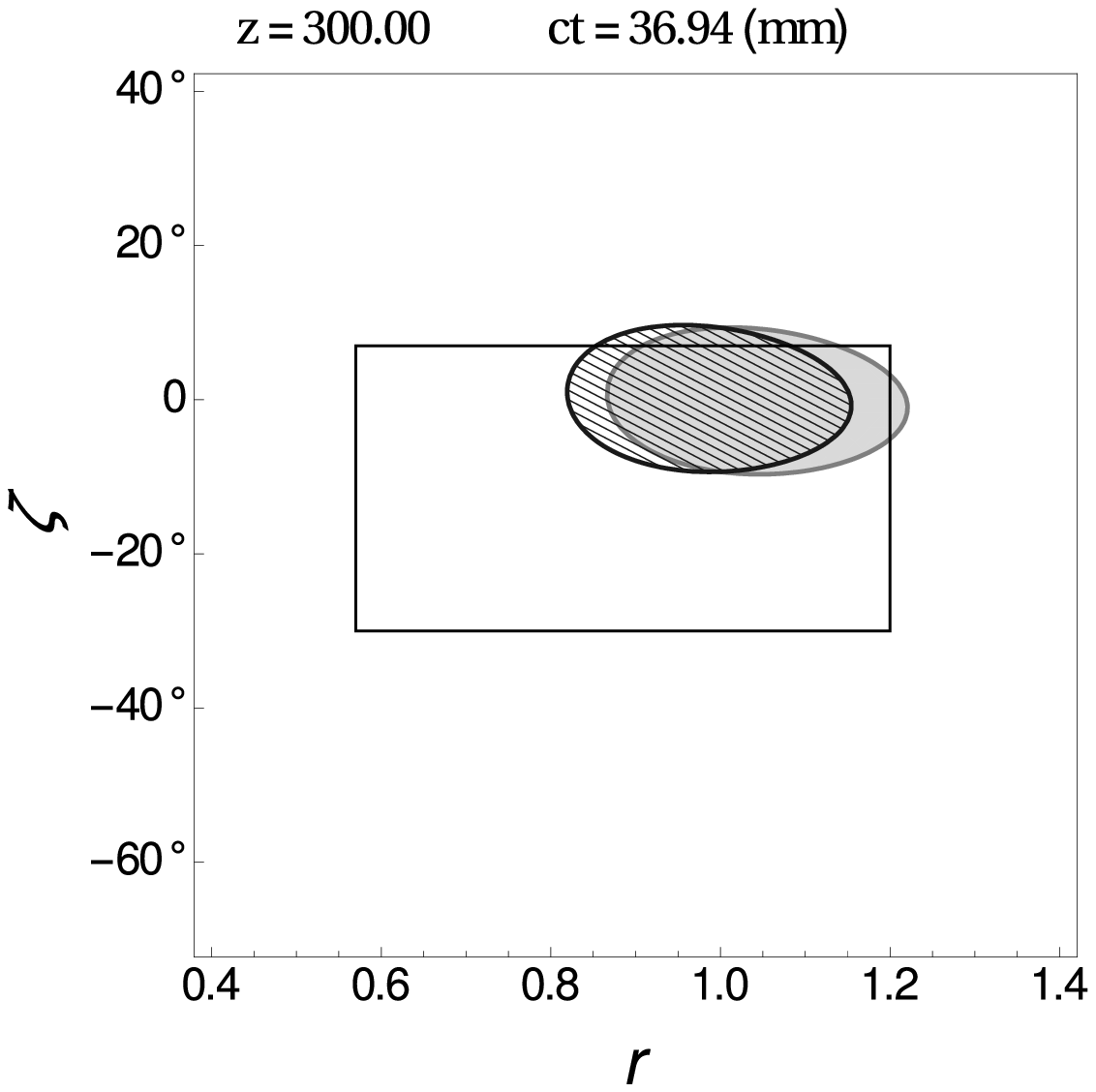,width=5.2cm}} 
	\end{tabular}
	\caption{\label{fig:rzeta_D} Areas of violation of the
      inequalities (\protect\ref{W-B-2-gammaT}) for $D$--mesons for
      the functions
      $\textrm{R}_{5,\, 6} (x,\, r,\,\zeta,\,\lambda)$ in the
      parameter plane of $r$ and $\zeta$ ($\zeta$ is measured in
      degrees) depending on $z$ or $ct$. The gray areas correspond
      to the violation of the function $\textrm{R}_{5} (x,\,
      r,\,\zeta,\,\lambda)$, while the hatched areas correspond to the
      function  $\textrm{R}_{6} (x,\,
      r,\,\zeta,\,\lambda)$. The experimentally allowed area of $r$
      and $\zeta$ is contained within the rectangle.}
\end{center}
\end{figure}

For neutral $D$--mesons the situation is very similar to the one with 
$K$--mesons. In order to study the violation of (\ref{W-B-2-gammaT}),
it is necessary to consider the dependence on $z$ of the functions $\textrm{R}_{5,\,
  6} (x,\, r,\,\zeta,\,\lambda)$. However, unlike the $K$--meson case, 
the parameters $r$ and $\zeta$ for $D$--mesons are not well fixed from
experiment. %
The evolution of the set of parameters $r$ and $\zeta$,  which violate
(\ref{W-B-2-gammaT}), with $z$ (or $ct$) is shown in FIG. \ref{fig:rzeta_D}. 
Gray areas correspond to the function $\textrm{R}_{5} (x,\,
r,\,\zeta,\,\lambda)$. Hatched areas correspond to 
$\textrm{R}_{6} (x,\, r,\,\zeta,\,\lambda)$.  The area of
experimentally allowed values of the parameters $r$ and $\zeta$ is
contained within the rectangle. At $t=0$ the areas do not intersect,
but have only a common point at $r =1$ and $\zeta = 0^o$. As $t \to
+\infty$ both areas shrink to the point $r =1$ and $\zeta = 0^o$,
corresponding to the results of Table II. 

\begin{figure}[p]
\begin{center}
	\begin{tabular}{cc}
	\mbox{\epsfig{file=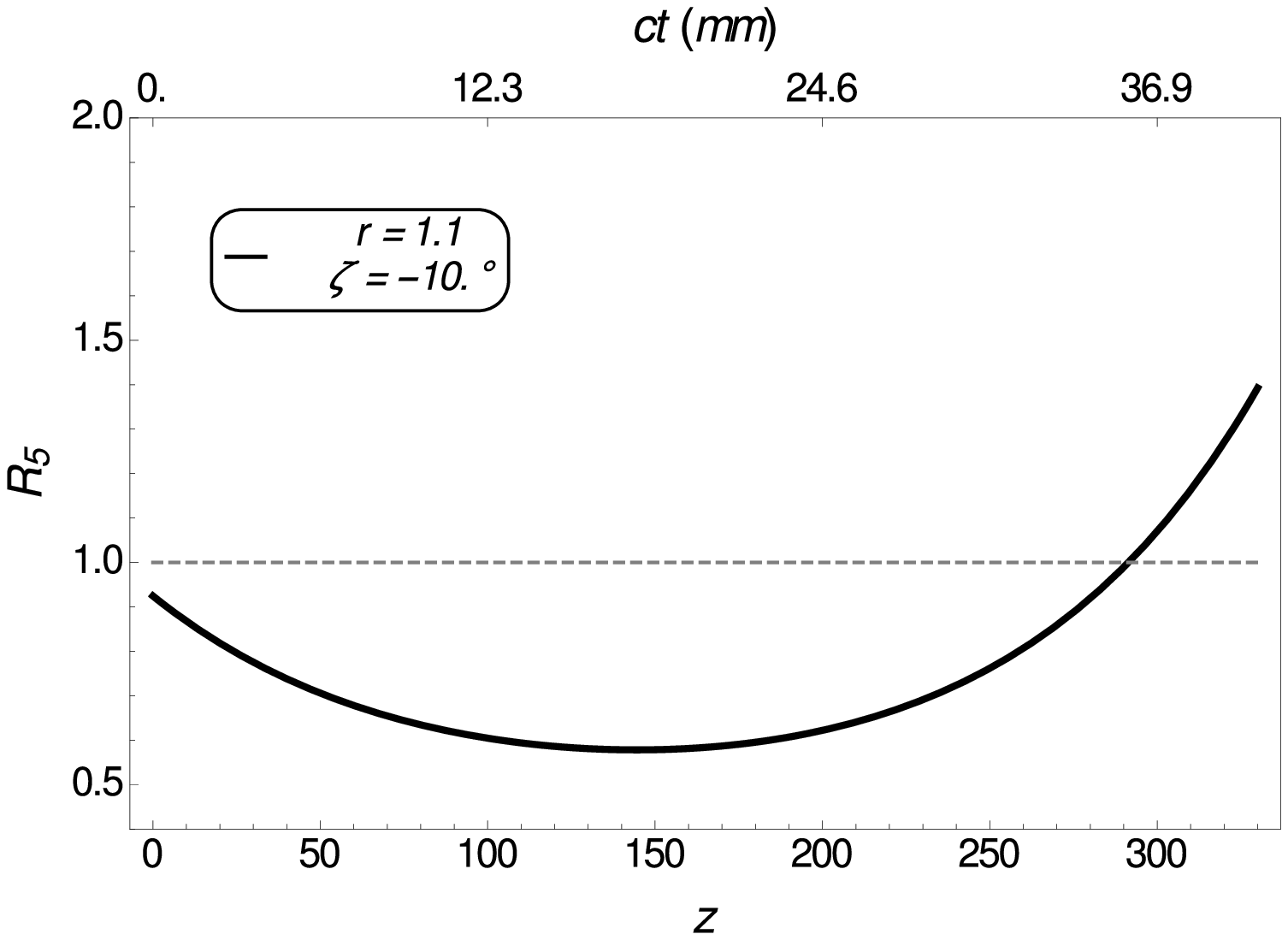,width=7.9cm}} & \mbox{\epsfig{file=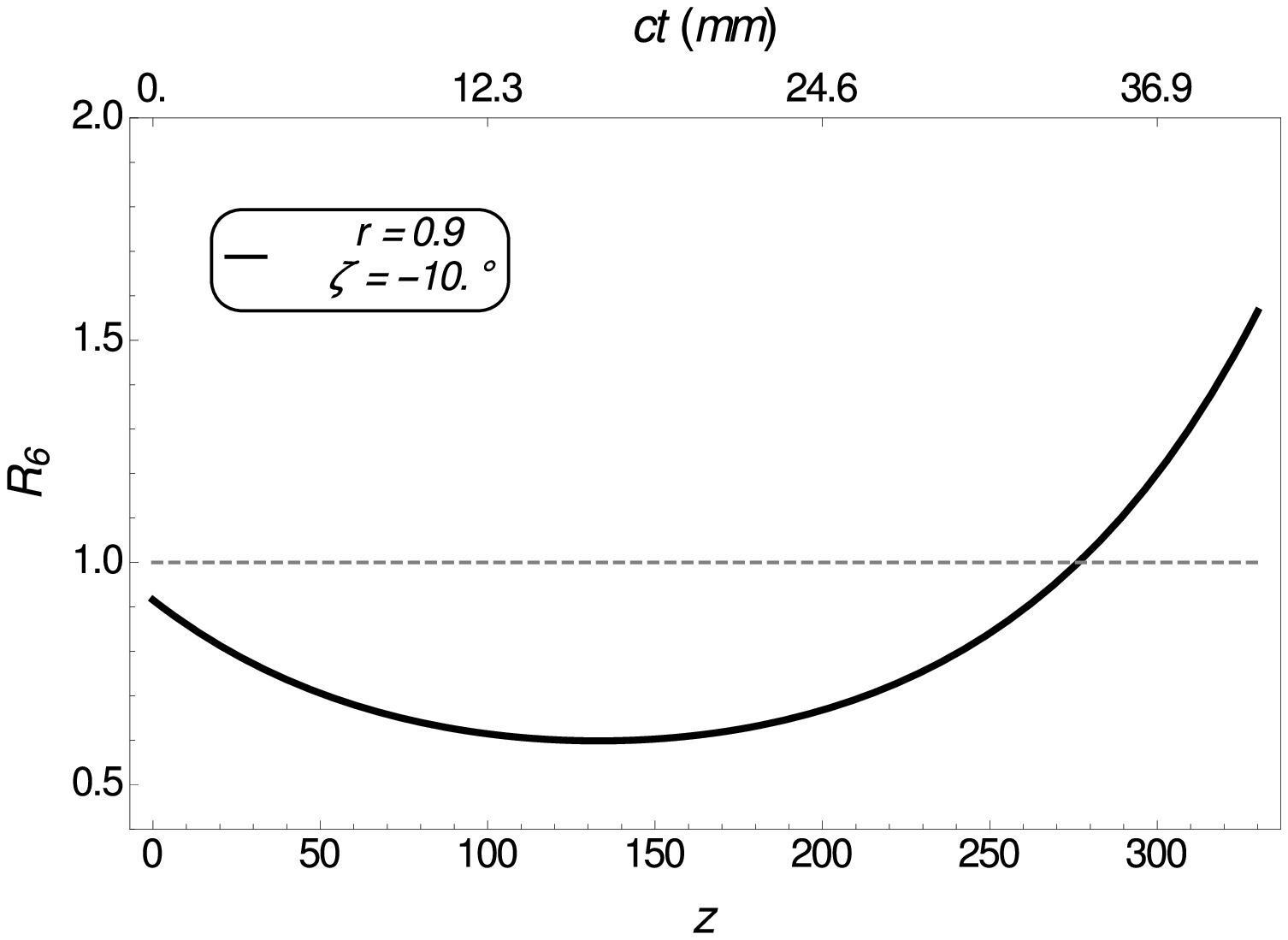,width=7.9cm}}
	\end{tabular}
	\caption{\label{fig:RHS_D} Functions $\textrm{R}_{5,\, 6} (x,\,
      r,\,\zeta,\,\lambda)$ for neutral $D$--mesons. The scale at the top
      corresponds to $c \,t$ (mm); the bottom scale corresponds to the
      time in units of the average lifetime $z = (\Gamma_H + \Gamma_L)\, t /2 =
      \Gamma\, t$, where $t$ is calculated in the $D$--meson rest
      frame. One can see that with the proper choice of the functions
      $\textrm{R}_{N}$ for $r >1$ and $r < 1$ the time-dependent
      Wigner inequalities (\protect\ref{W-B-2-gammaT}) are violated in
      the whole experimentally accessible range of $z$.
}
\end{center}
\end{figure}

\begin{figure}[p]
\begin{center}
	\begin{tabular}{cc}
	\mbox{\epsfig{file=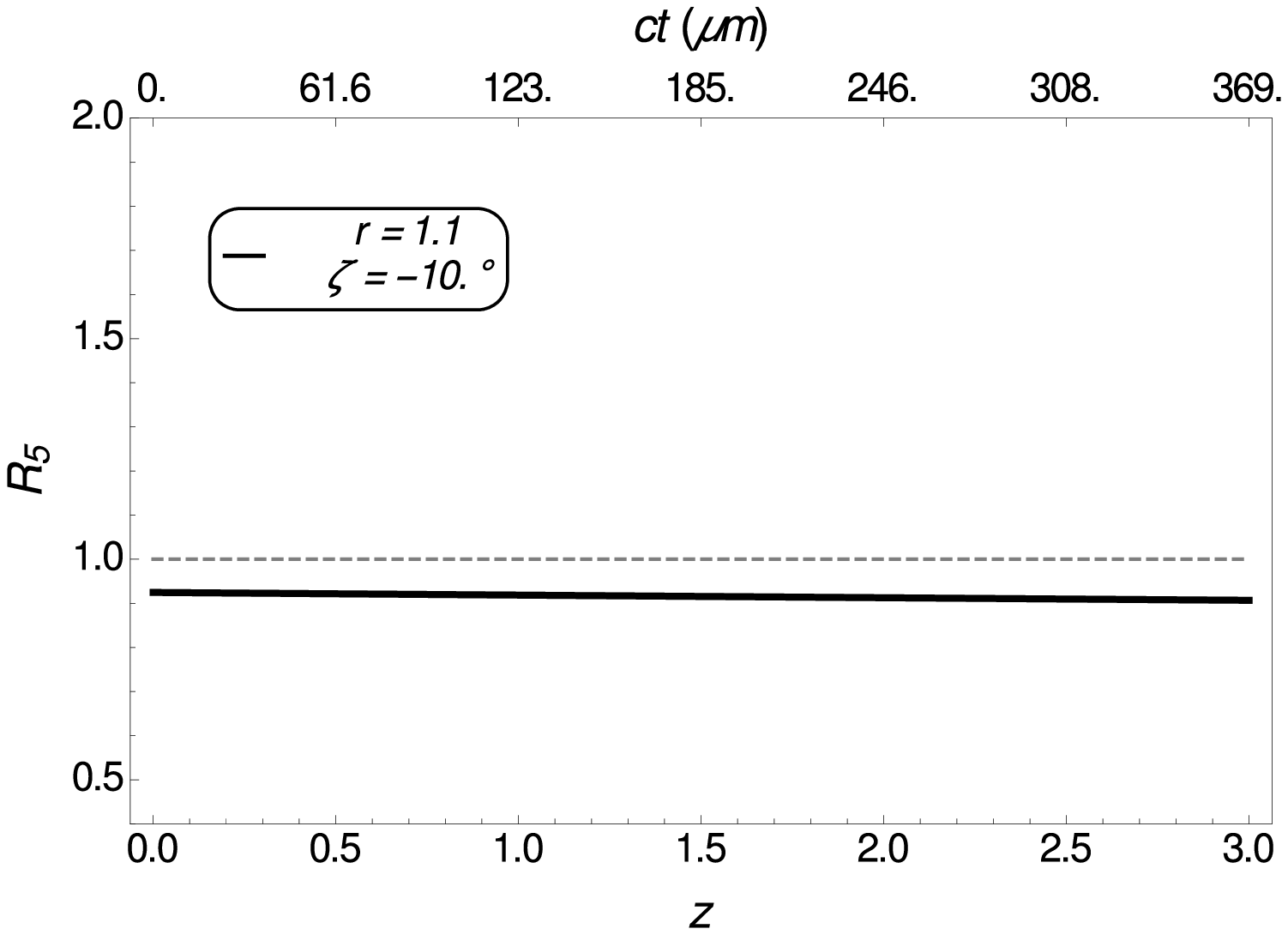,width=7.9cm}} & \mbox{\epsfig{file=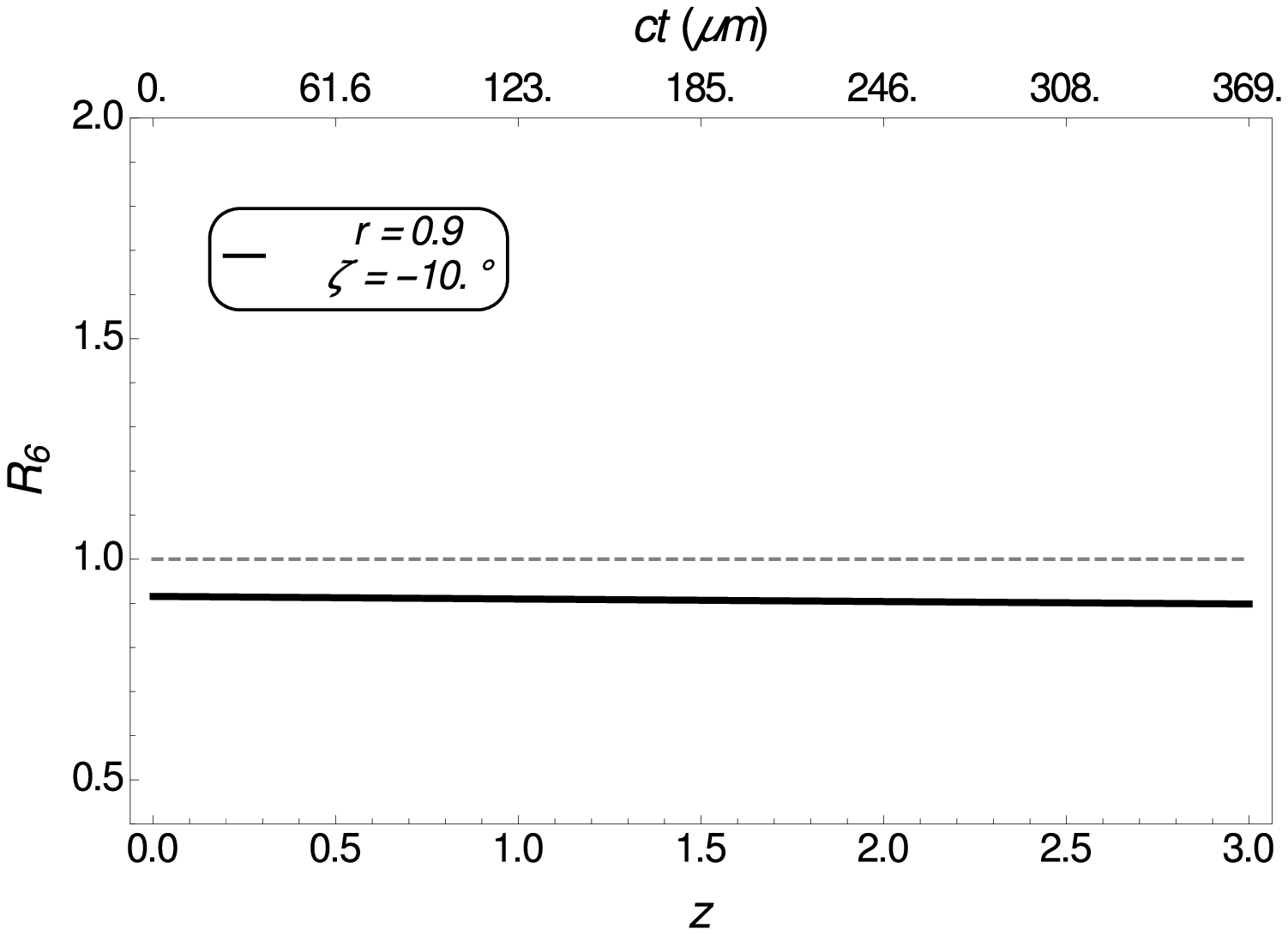,width=7.9cm}}
	\end{tabular}
	\caption{\label{fig:RHS_D3} Functions $\textrm{R}_{5,\, 6} (x,\,
      r,\,\zeta,\,\lambda)$ for neutral $D$--mesons in the area $z
      \le 3$. The  $c t$ here is measured in microns ($\mu$m).}
\end{center}
\end{figure}

From FIG. \ref{fig:rzeta_D} one can see that with $r > 1$ in the limit $t \to
0$, inequality (\ref{W-B-2-gammaT}) is only violated for the
function $\textrm{R}_{5} (x,\, r,\,\zeta,\,\lambda)$. For $r < 1$ in the limit $t \to
0$ the violation only happens for the function $\textrm{R}_{6}
(x,\, r,\,\zeta,\,\lambda)$. This statement is illustrated in
FIG. \ref{fig:RHS_D}. For the function $\textrm{R}_{5}
(x,\, r,\,\zeta,\,\lambda)$, $r$ is set to $1.1$. For the function
$\textrm{R}_{6} (x,\, r,\,\zeta,\,\lambda)$, $r$ is set to  $0.9$. The
value of $\zeta$ in both cases is set to $-10^o$. 
In analogy with the case of $K$--mesons, there is a restoration of the inequalities
(\ref{W-B-2-gammaT}), but with a higher value of $z$, $\sim 280$,
which is beyond experimental reach.
Maximal violation of (\ref{W-B-2-gammaT}) also happens at $z\sim
150$. From FIG. \ref{fig:RHS_D3} one can see that in the
experimentally allowed area $z \le3$ the violation of (\ref{W-B-2-gammaT})
is less than 10\%.

The difference in behaviour of the functions $\textrm{R}_{5,\, 6} (x,\,
r,\,\zeta,\,\lambda)$ for $K$-- and $D$--mesons is linked to the value
of the ratio $|\Delta \Gamma | / \Gamma$, which sets the scale of the
horizontal axis. For $K$--mesons, $\displaystyle
\left (\frac{|\Delta \Gamma |}{\Gamma} \right )_K \approx 2$, while for
$D$--mesons this parameter is smaller by almost two orders of
magnitude, $\displaystyle \left (\frac{|\Delta \Gamma |}{\Gamma} \right
)_D \approx 10^{-2}$.

\begin{figure}[p]
\begin{center}
	\begin{tabular}{ccc}
	\mbox{\epsfig{file= 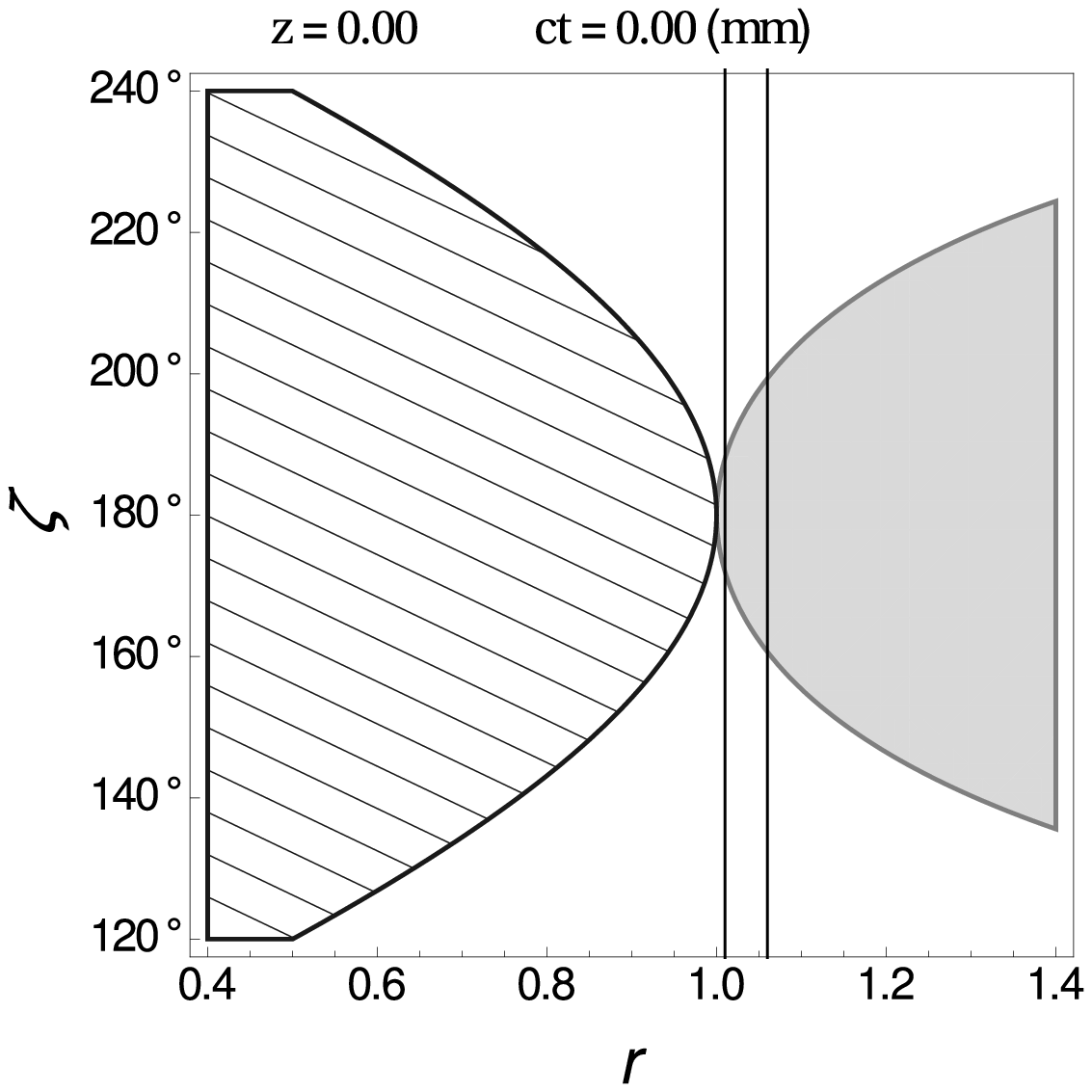,width=5.2cm}} & \mbox{\epsfig{file=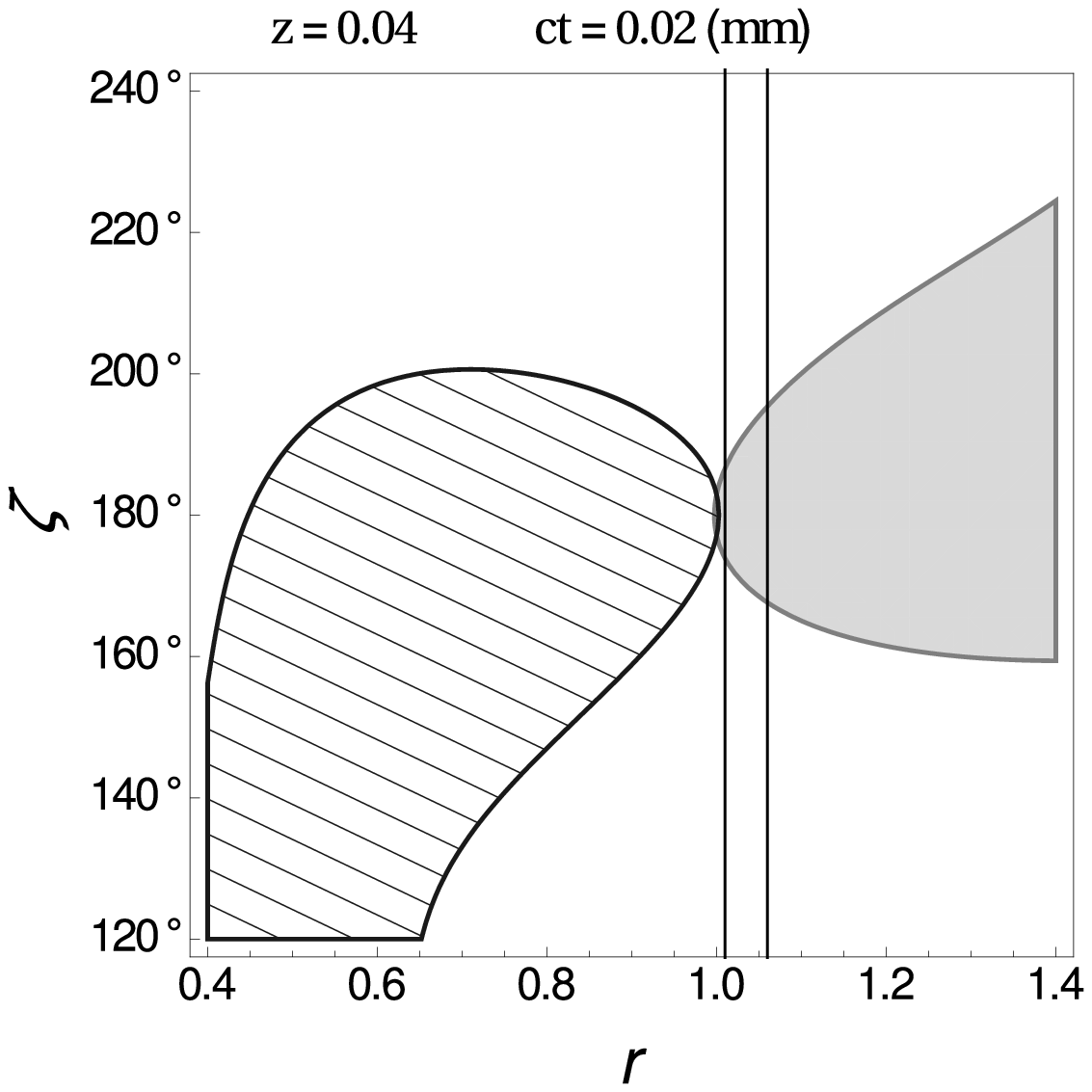,width=5.2cm}} & \mbox{\epsfig{file=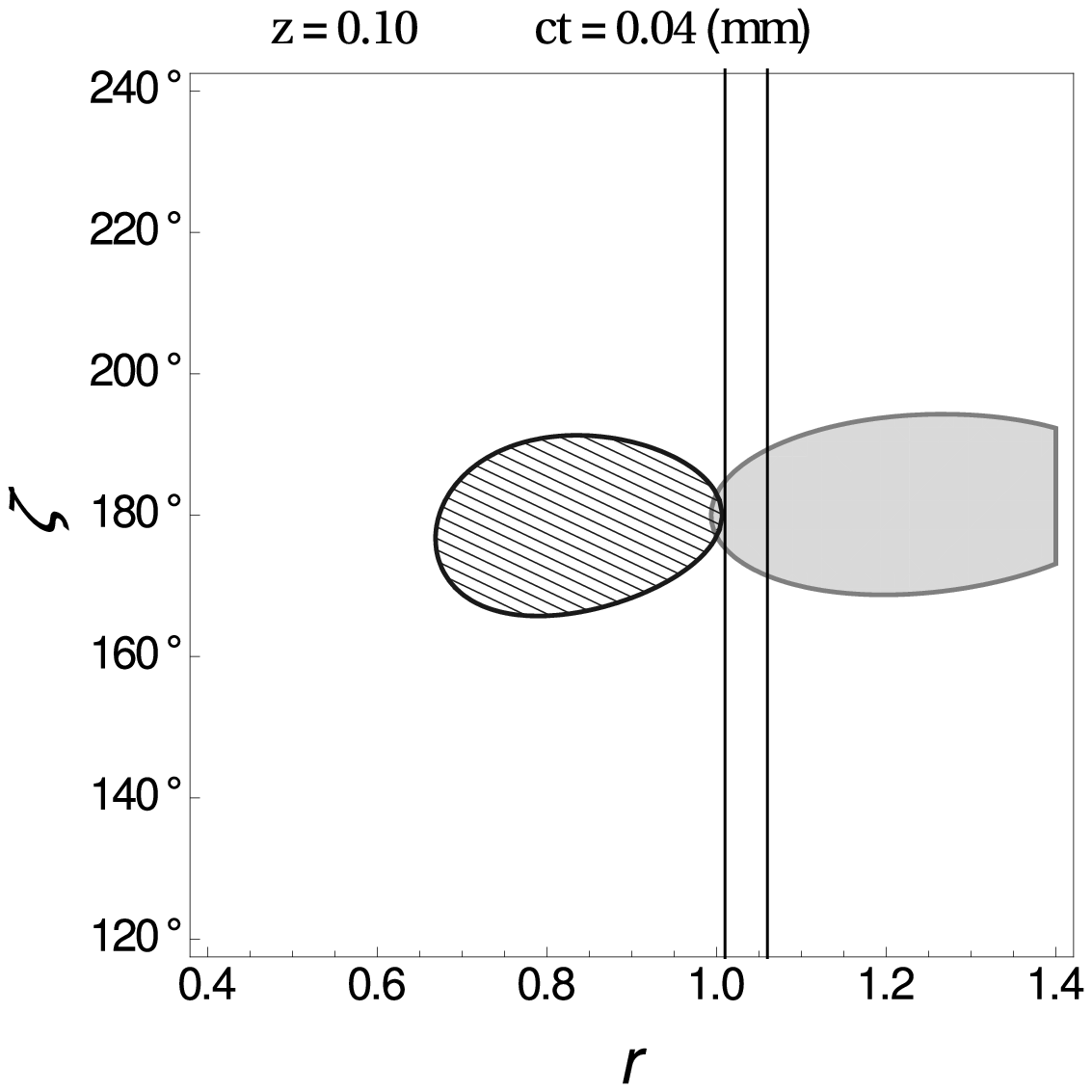,width=5.2cm}} \\
         \mbox{\epsfig{file=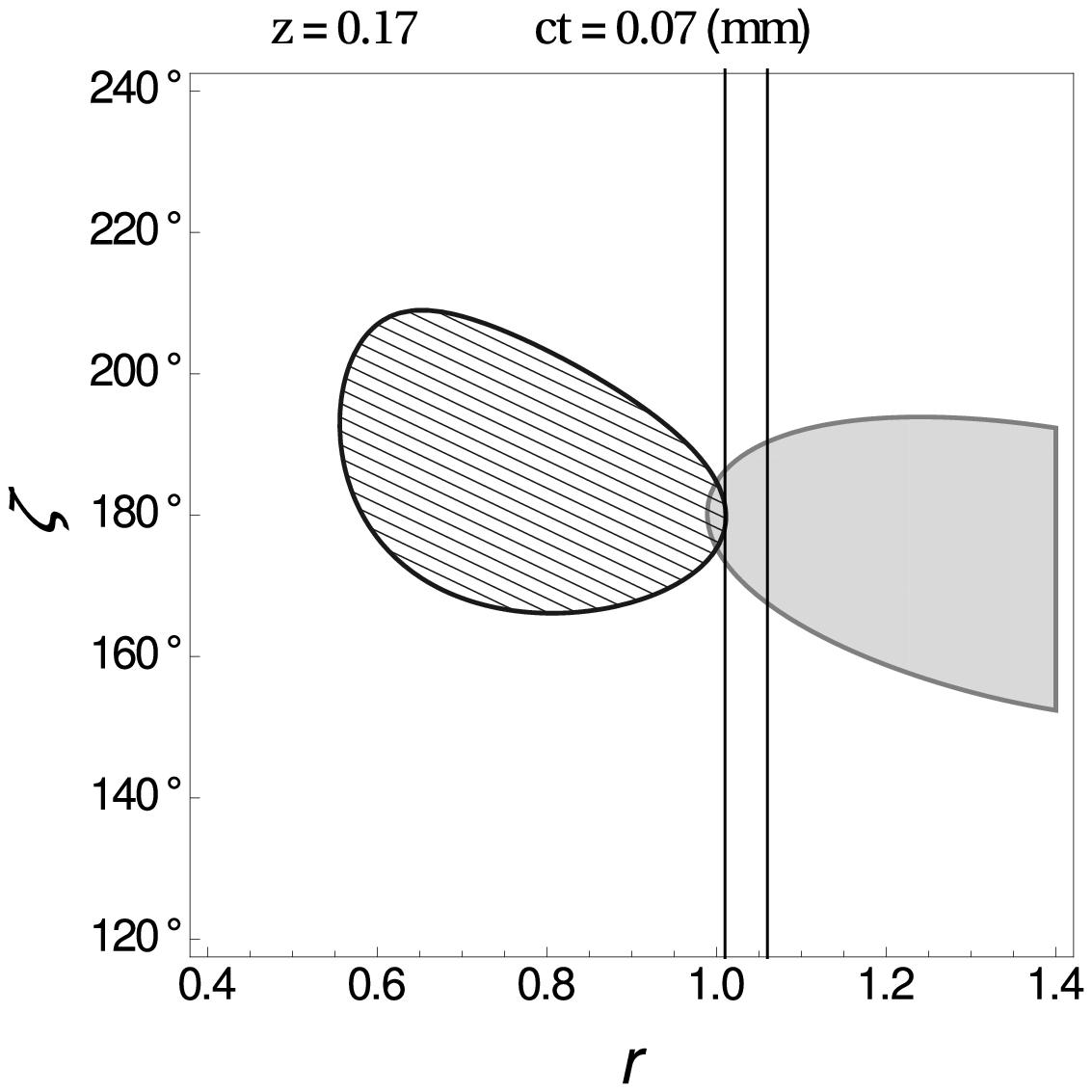,width=5.2cm}} & \mbox{\epsfig{file=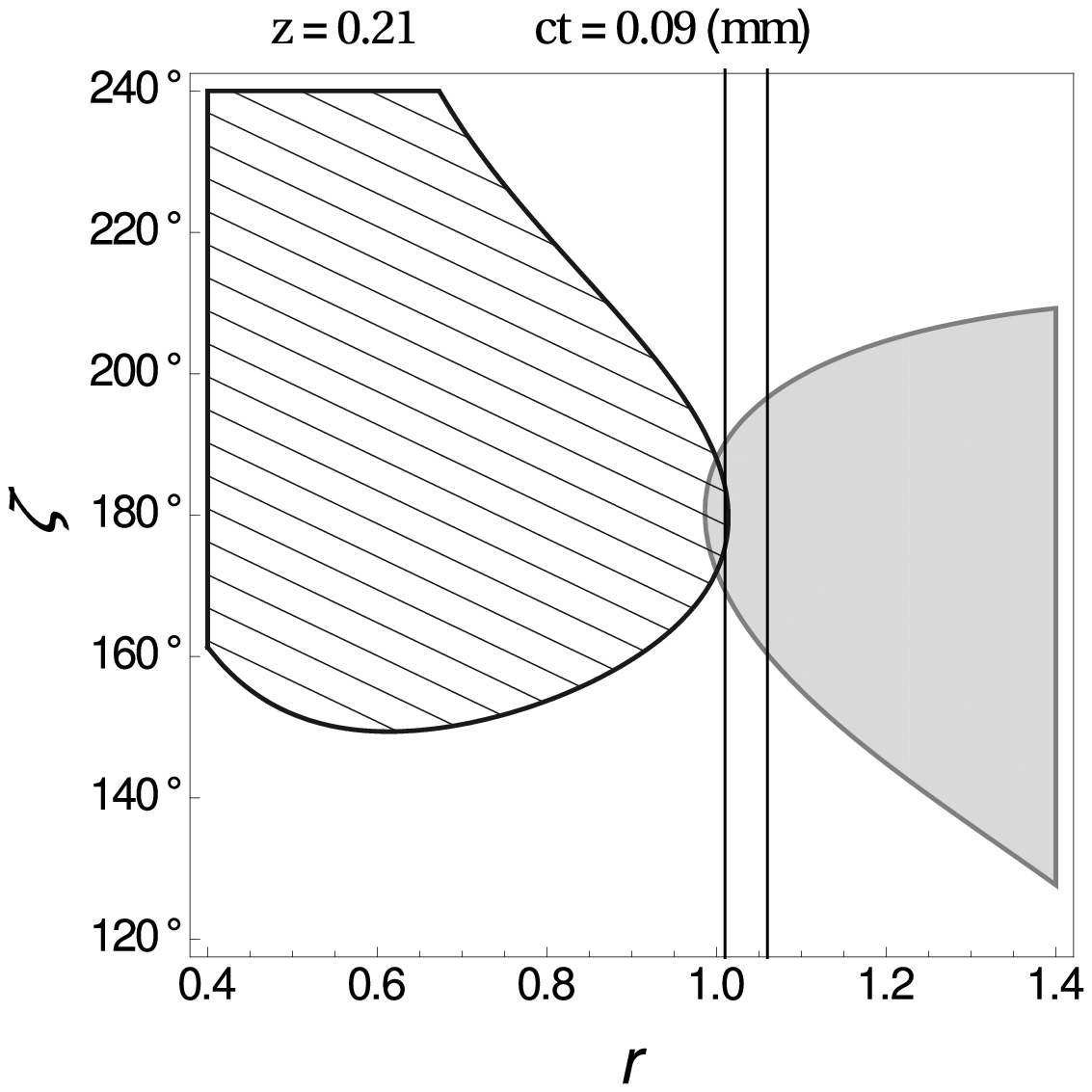,width=5.2cm}} & \mbox{\epsfig{file=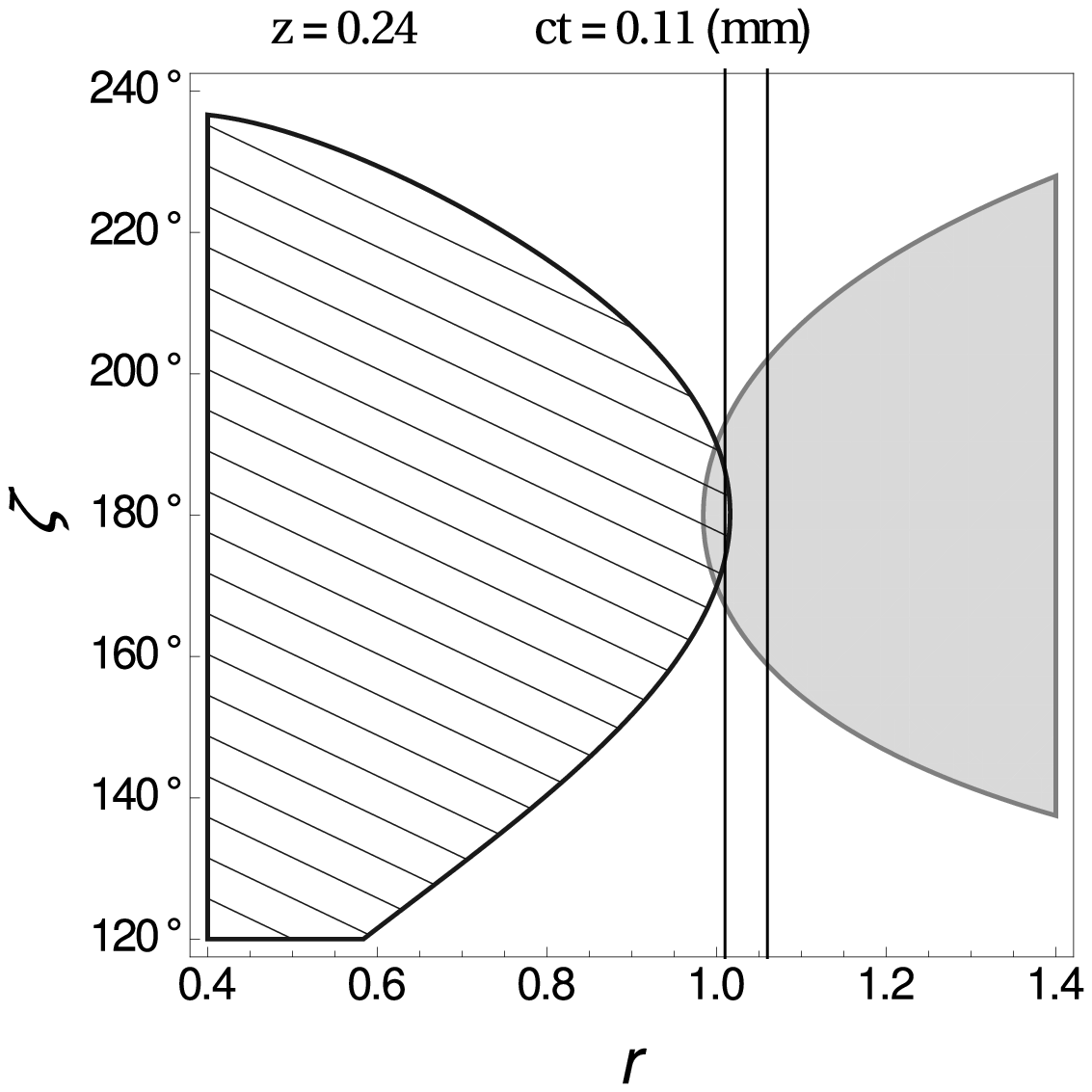,width=5.2cm}} \\
         \mbox{\epsfig{file=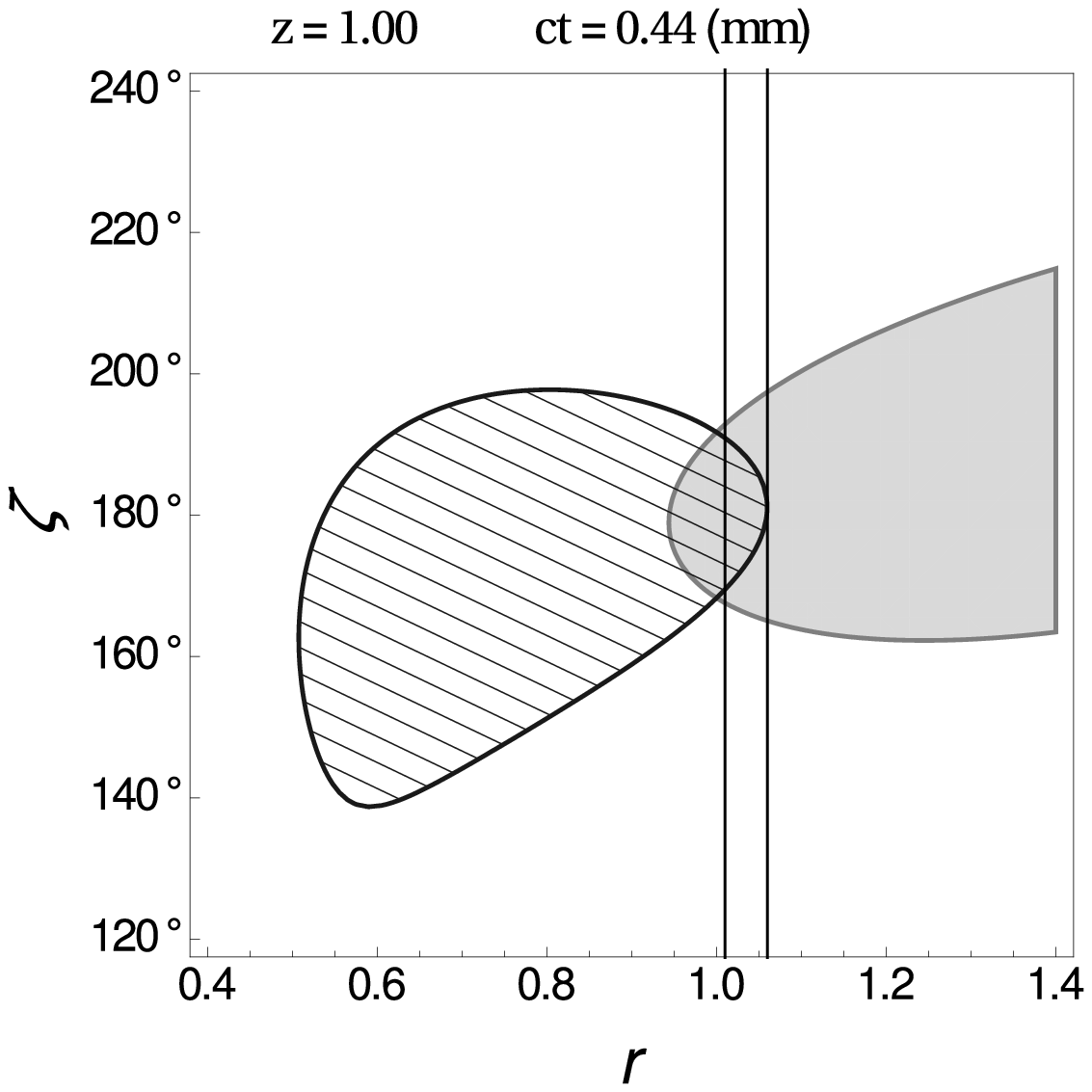,width=5.2cm}} & \mbox{\epsfig{file=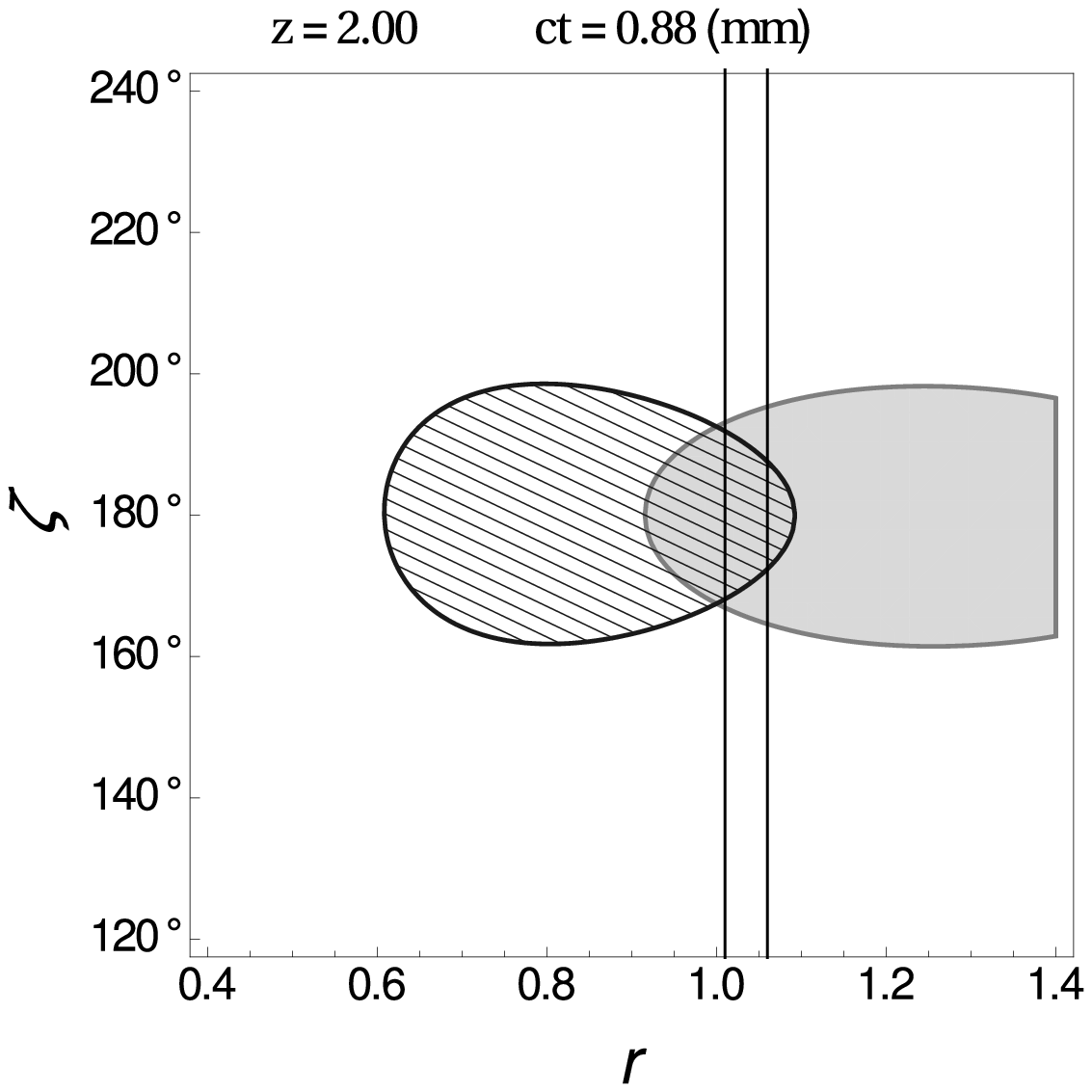,width=5.2cm}} & \mbox{\epsfig{file=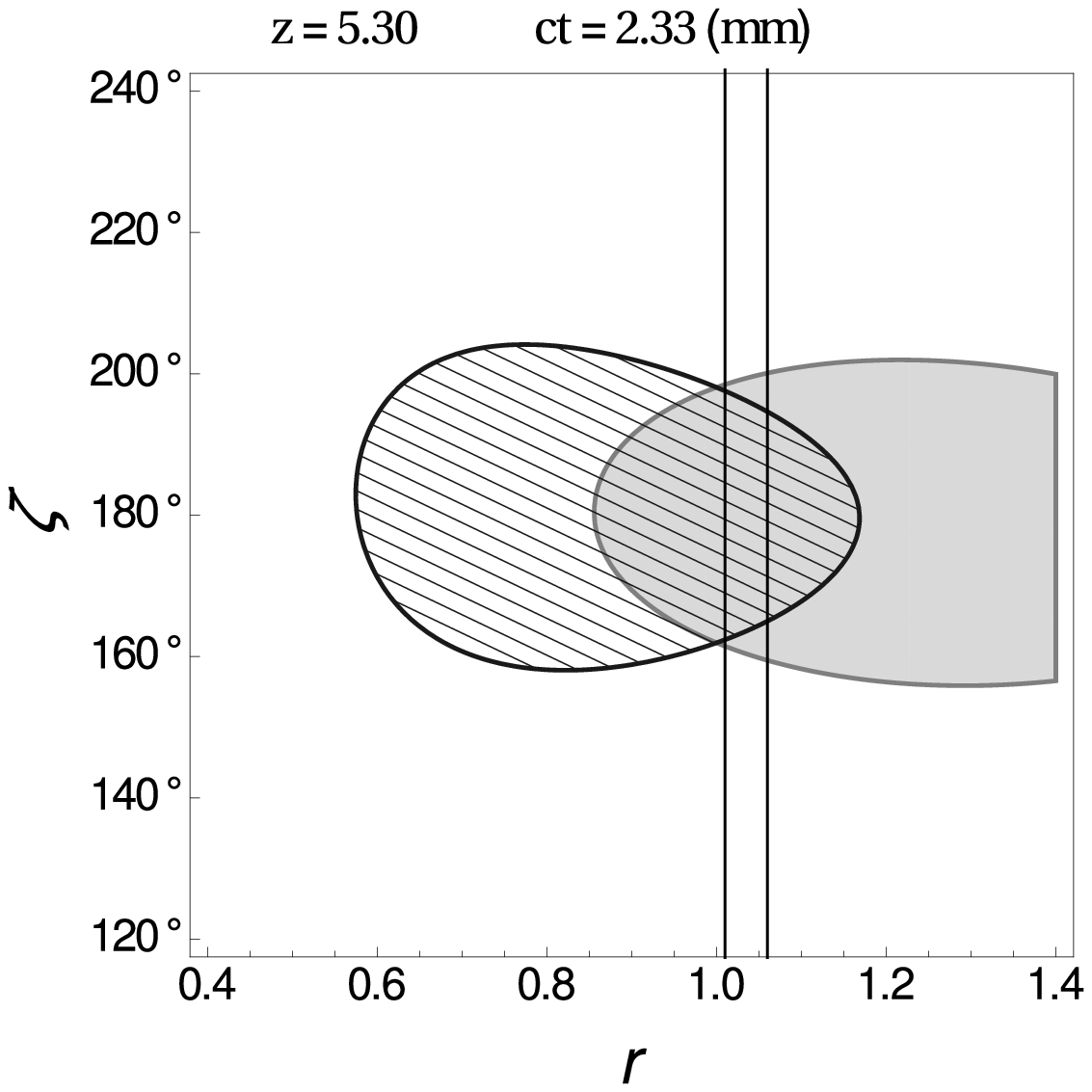,width=5.2cm}} 
	\end{tabular}
	\caption{\label{fig:rzeta_Bs} Areas of the violation of
      (\protect\ref{W-B-2-gammaT}) for  
      $B_s$--mesons for the functions $\textrm{R}_{7,\, 8} (x,\,
      r,\,\zeta,\,\lambda)$ in the $r$--$\zeta$ plane
      ($\zeta$ is measured in degrees). The gray areas
      correspond to the function $\textrm{R}_{7} (x,\, r,\,\zeta,\,\lambda)$
      the hatched areas correspond to the function $\textrm{R}_{8} (x,\,
      r,\,\zeta,\,\lambda)$. The vertical band corresponds to the
      experimentally allowed area of $r$ and $\zeta$.}
\end{center}
\end{figure}

As was pointed out above, the study of the
violation of (\ref{W-B-2-gammaT}) for $B_s$--meson systems requires
the functions $\textrm{R}_{7,\, 8} (x,\, r,\,\zeta,\,\lambda)$. In FIG.
\ref{fig:rzeta_Bs} we show how the areas of violation of
(\ref{W-B-2-gammaT}) depend on $z$ or $ct$ for the functions
$\textrm{R}_{7} (x,\, r,\,\zeta,\,\lambda)$ (gray area) and
$\textrm{R}_{8} (x,\, r,\,\zeta,\,\lambda)$ (hatched area).
The vertical band shows the experimentally allowed values of $r$ and
$\zeta$. For $t=0$ the areas have a single point of intersection,
$r=1$ and $\zeta = 180^o$. As $t \to +\infty$ they shrink to a point
at $(1,\, 180^o)$. Unlike the case for $D$--mesons, the areas of violation for the
$B_s$--mesons do not evolve monotonical. This is due to the
oscillations of $B_s$--mesons, which play an important role here.

It is experimentally established that for $B_s$--mesons, $r > 1$. Hence for 
$z \to 0$ only the function $\textrm{R}_{7} (x,\, r,\,\zeta,\,\lambda)$
violates the inequality (\ref{W-B-2-gammaT}). However in  FIG. 
\ref{fig:RHS_Bs3} one observes that for $z \gtrsim  \Gamma / (2 \Delta M )
\sim 1/2$ the function  $\textrm{R}_{8} (x,\, r,\,\zeta,\,\lambda)$
also begins to violate (\ref{W-B-2-gammaT}). For the numerical
simulation, the following values of the parameters were used: $r =
1.004$ and $\zeta = 185^o$. The maximum  violation of
(\ref{W-B-2-gammaT}) is reached in the area $z \sim 20$. At $z
\sim 40$ the inequalities are not violated, as in $D$--meson
systems.  Due to the high value of $z$ this effect is not
experimentally reachable. As $\displaystyle \left (\frac{|\Delta \Gamma
    |}{\Gamma} \right )_{B_z} \approx 0.13$, the corresponding values
of  $z$ are intermediate  between the ones for $K$-- and $D$--mesons.

\begin{figure}[p]
\begin{center}
	\begin{tabular}{cc}
	\mbox{\epsfig{file=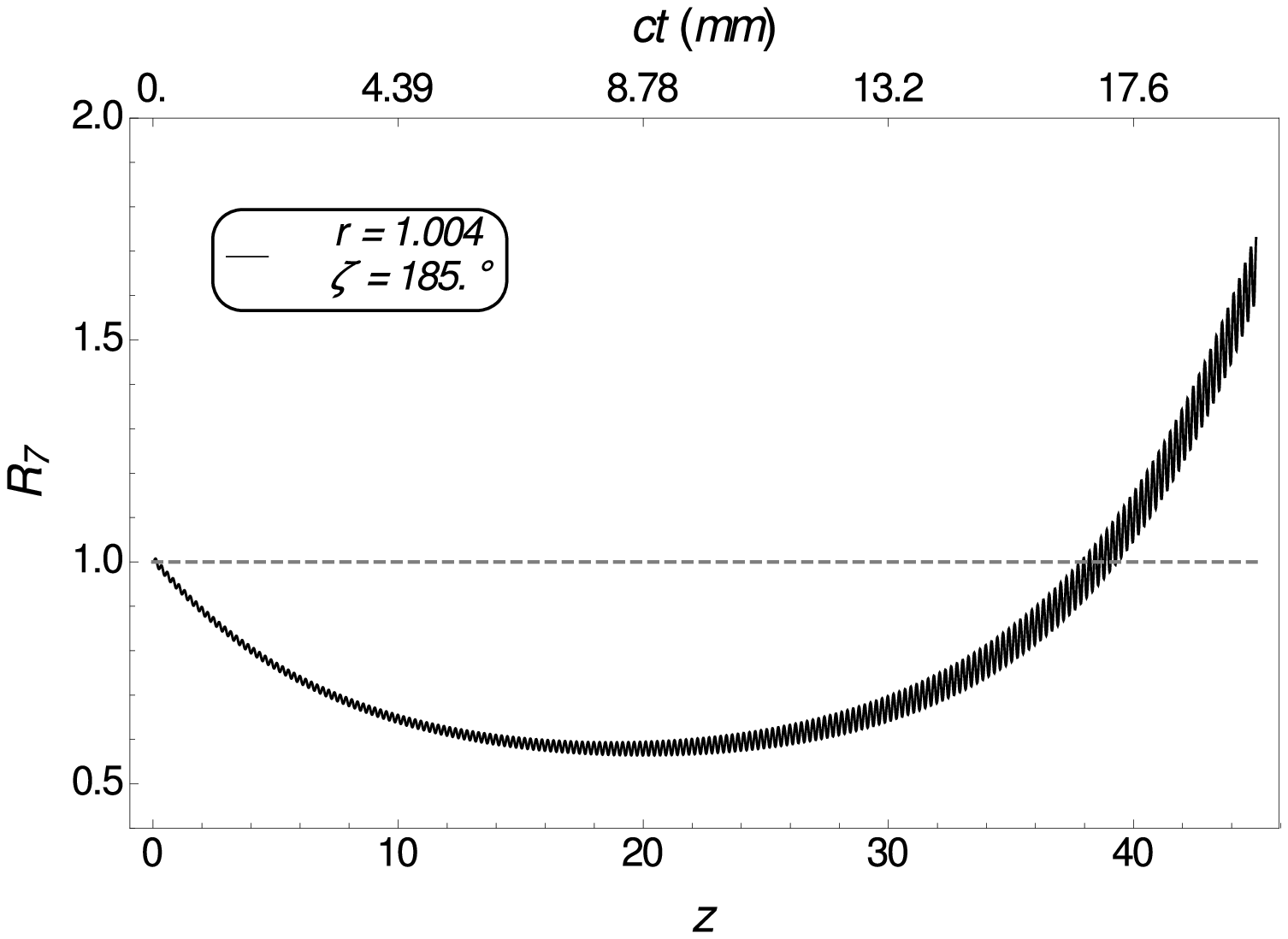,width=7.9cm}} & \mbox{\epsfig{file=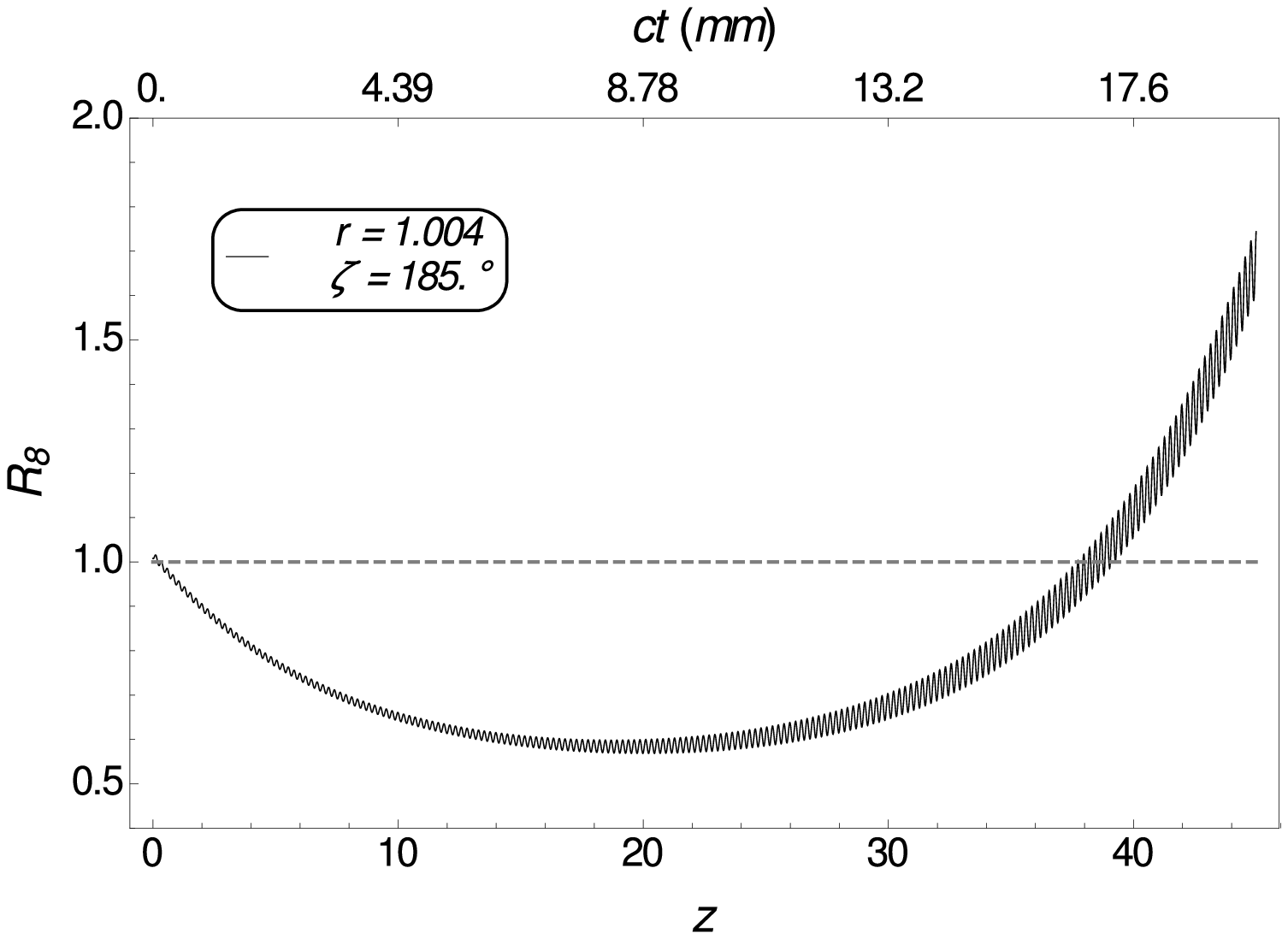,width=7.9cm}}
	\end{tabular}
	\caption{\label{fig:RHS_Bs} Functions $\textrm{R}_{7,\, 8} (x,\,
      r,\,\zeta,\,\lambda)$ for $B_s$--mesons. The scale at the top
      corresponds to the $c \,t$ (mm), while  the bottom scale corresponds to
      time in units of the average lifetime $z = (\Gamma_H + \Gamma_L)\, t /2 =
      \Gamma\, t$, where $t$ is calculated in the 
      $B_s$--meson rest frame. One can see that the time-dependent inequalities
      (\protect\ref{W-B-2-gammaT})) are violated (taking the proper
      $\textrm{R}_{N}$ for $r >1$ and $r < 1$) in almost all of the
      experimentally allowed range of $z$.
}
\end{center}
\end{figure}

\begin{figure}[p]
\begin{center}
	\begin{tabular}{cc}
	\mbox{\epsfig{file=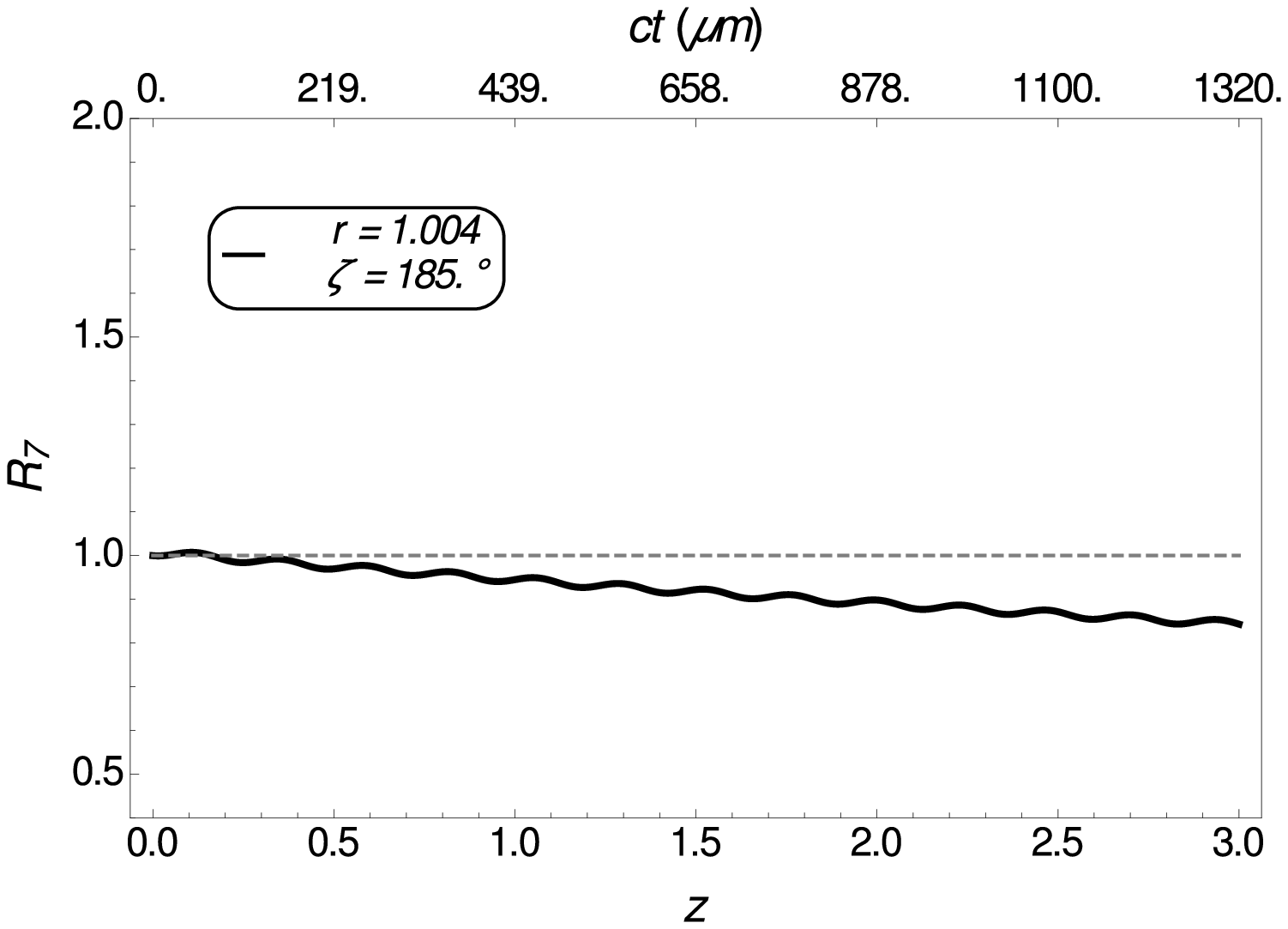,width=7.9cm}} & \mbox{\epsfig{file=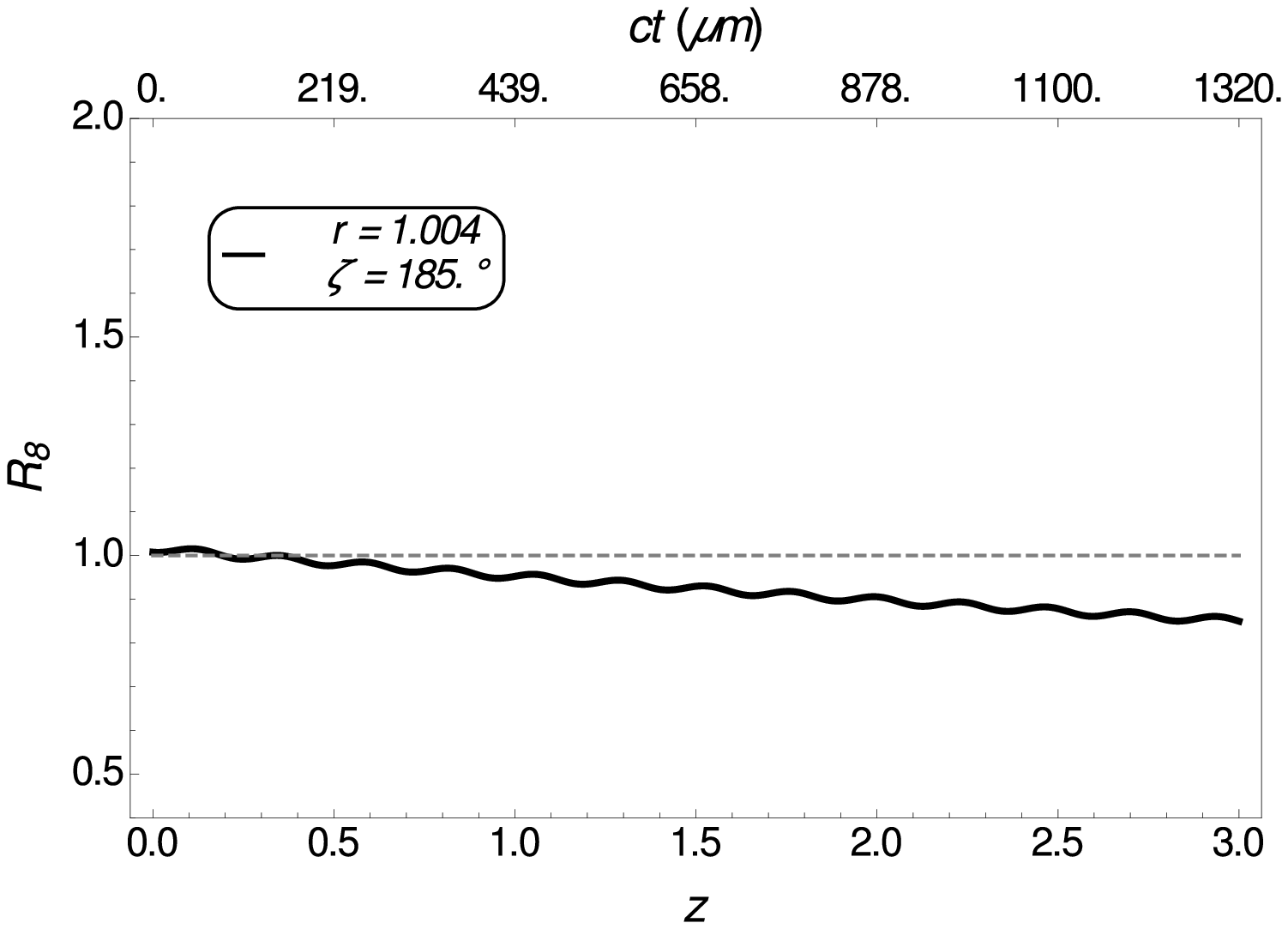,width=7.9cm}}
	\end{tabular}
	\caption{\label{fig:RHS_Bs3} Functions $\textrm{R}_{7,\,8} (x,\,
      r,\,\zeta,\,\lambda)$ for $B_s$--mesons in the range $z \le 3$,
      which is the most experimentally interesting.
      Both functions are almost consistent with one, while the 
      $\textrm{R}_{8}$ slightly exceeds one at $z \to 0$. Unlike 
      FIG. \protect\ref{fig:RHS_Bs}, here the $c t$ is in microns ($\mu$m).}
\end{center}
\end{figure}

\section*{Conclusions}

Using the oscillations of neutral pseudoscalar mesons we demonstrate
the advantages of the time-dependent Wigner inequality (\ref{W-B-4}) over the
static inequality (\ref{W-B-2}). Eight new time-dependent inequalities
(\ref{W-B-2-gammaT}) were obtained. They can be violated by proper
choices of $\Delta\Gamma$ and $\nicefrac{q}{p}$ for  
$K$--, $D$-- and $B_s$--mesons. Relaxation of the
obtained inequalities at high values of the variable $z$, is
found. This effect is governed explicitly by the $CP$--violation
parameters of the considered systems. The inequalities
(\ref{W-B-2-gammaT}) may be tested at contemporary high-energy
physics experiments

\section{Acknowledgements}
The authors would like to express deep gratitude to Dr. S.~Baranov 
(Lebedev Physical Institute, Russia) for numerous discussions related 
to tests of the Bell inequalities in particle physics; to 
Dr. A.~Grinbaum (CEA-Saclay, SPEC/LARSIM, France) for educational 
chats on the foundations of quantum theory; to Prof. S. Seidel 
(University of New Mexico, USA) for help with preparation of the 
paper. %

We thank for support the Russian Ministry of Education and Science
(grant  N14.610.21.0002, ID RFMEFI61014X0002)  and the Russian program
of support of the leading scientific schools (grant SS-3042.2014.2).

\appendix

\section{Probabilities required for the time-dependent Wigner inequalities}
\label{sec:A}

In this appendix we summarize all the probabilities that are required to
obtain the static (\ref{W-B-2})  and time-dependent (\ref{W-B-4}) Wigner
inequalities for the correlated systems of neutral
pseudoscalar mesons.

In the framework of quantum theory, using the normalization condition 
(\ref{pq-normirovka}) and the initial condition 
(\ref{correlationBbarB-t=0}), one can obtain the following expressions
for the time-independent probabilities:

\begin{eqnarray}
\label{w-BbarB-I}
&& w(M_1^{(2)},\, \bar M^{(1)},\, t_0)\,=\,  \left |\bra{M_1^{(2)}}\bracket{\bar M^{(1)}}{\Psi (t_0)}\right |^2\, =\,\frac{1}{4}\,\equiv
\, \frac{1}{4}\,\left ( |p|^2 + |q|^2 \right );\nonumber \\
&& w(M_1^{(2)},\, M^{(1)},\, t_0)\,=\,  \left |\bra{M_1^{(2)}}\bracket{M^{(1)}}{\Psi (t_0)}\right |^2\, =\,\frac{1}{4}\,\equiv
\,\frac{1}{4}\,\left ( |p|^2 + |q|^2 \right );\nonumber \\
&& w(M_2^{(2)},\, \bar M^{(1)},\, t_0)\,=\,  \left |\bra{M_2^{(2)}}\bracket{\bar M^{(1)}}{\Psi (t_0)}\right |^2\, =\,\frac{1}{4}\,\equiv
\, \frac{1}{4}\,\left ( |p|^2 + |q|^2 \right );\nonumber \\
&& w(M_2^{(2)},\, M^{(1)},\, t_0)\,=\,  \left |\bra{M_2^{(2)}}\bracket{M^{(1)}}{\Psi (t_0)}\right |^2\, =\,\frac{1}{4}\,\equiv
\,\frac{1}{4}\,\left ( |p|^2 + |q|^2 \right );\\
&& w(M_1^{(2)},\, M_H^{(1)},\, t_0)\,=\, \left |\bra{M_1^{(2)}}\bracket{M^{(1)}_H}{\Psi (t_0)}\right |^2\, =\,
\frac{1}{4}\, \left |p + q\right |^2 = \frac{1}{4} \left ( 1\, +\,\cos\zeta\,\sin ( 2 \beta )\right); \nonumber \\
&& w(M_2^{(2)},\, M_H^{(1)},\, t_0)\,=\, \left |\bra{M_2^{(2)}}\bracket{M^{(1)}_H}{\Psi (t_0)}\right |^2\, =\,
\frac{1}{4}\, \left | p - q\right |^2 = \frac{1}{4} \left ( 1\, -\,\cos\zeta\,\sin ( 2 \beta ) \right); \nonumber \\
&& w(M_1^{(2)},\, M_L^{(1)},\, t_0)\,=\, \left |\bra{M_1^{(2)}}\bracket{M^{(1)}_L}{\Psi (t_0)}\right |^2\, =\,
\frac{1}{4}\, \left |p - q\right |^2 = \frac{1}{4} \left ( 1\, -\,\cos\zeta\,\sin ( 2 \beta )\right); \nonumber \\
&& w(M_2^{(2)},\, M_L^{(1)},\, t_0)\,=\, \left |\bra{M_2^{(2)}}\bracket{M^{(1)}_L}{\Psi (t_0)}\right |^2\, =\,
\frac{1}{4}\, \left | p + q\right |^2 = \frac{1}{4} \left ( 1\, +\,\cos\zeta\,\sin ( 2 \beta ) \right); \nonumber \\
&& w(M_H^{(2)},\, \bar M^{(1)},\, t_0)\,=\,\left |\bra{M_H^{(2)}}\bracket{\bar M^{(1)}}{\Psi (t_0)}\right |^2\, =\,
\frac{1}{2}\, \left | p \right |^2 =  \frac{1}{2}\, \cos^2 \beta; \nonumber \\
&& w(M_H^{(2)},\, M^{(1)},\, t_0)\,=\,\left |\bra{M_H^{(2)}}\bracket{M^{(1)}}{\Psi (t_0)}\right |^2\, =\,
\frac{1}{2}\, \left | q \right |^2 = \frac{1}{2}\, \sin^2\beta; \nonumber \\
&& w(M_L^{(2)},\, \bar M^{(1)},\, t_0)\,=\,\left |\bra{M_L^{(2)}}\bracket{\bar M^{(1)}}{\Psi (t_0)}\right |^2\, =\,
\frac{1}{2}\, \left | p \right |^2 =  \frac{1}{2}\, \cos^2 \beta; \nonumber \\
&& w(M_L^{(2)},\, M^{(1)},\, t_0)\,=\,\left |\bra{M_L^{(2)}}\bracket{M^{(1)}}{\Psi (t_0)}\right |^2\, =\,
\frac{1}{2}\, \left | q \right |^2 = \frac{1}{2}\, \sin^2\beta. \nonumber 
\end{eqnarray}

To obtain the inequality (\ref{W-B-4}) for correlated pairs
of mesons $M \bar M$, we need the values of the following
time-dependent probabilities ($t_0 = 0$ below):
\begin{eqnarray}
\label{w-BbarB-II}
&& w(M_1(0) \to M_1(t)) =  \left |\bracket{M_1(t)}{M_1} \right |^2= 
      \left | 
              g_+(t)\, -\, \frac{1}{2}\,\left ( \frac{q}{p} + \frac{p}{q}\right )\, g_-(t)
     \right |^2; \nonumber \\
&& w(M_2 (0) \to M_1 (t)) =  \left |\bracket{M_1(t)}{M_2} \right |^2=  \left | \frac{1}{2}\,\left ( \frac{q}{p} - \frac{p}{q}\right )\ g_- (t) \right |^2; \nonumber \\
&& w(M_2(0) \to M_2(t)) =  \left |\bracket{M_2(t)}{M_2} \right |^2= 
      \left | 
              g_+(t)\, +\, \frac{1}{2}\,\left ( \frac{q}{p} + \frac{p}{q}\right )\, g_-(t)
     \right |^2; \nonumber \\
&& w(M_1 (0) \to M_2 (t)) =  \left |\bracket{M_2(t)}{M_1} \right |^2=  \left | \frac{1}{2}\,\left ( \frac{q}{p} - \frac{p}{q}\right )\ g_- (t) \right |^2; \nonumber \\
&& w(\bar M (0) \to \bar M (t)) =  \left |\bracket{\bar M (t)}{\bar M} \right |^2 =  |g_+ (t)|^2; \\
&& w(M (0) \to \bar M (t)) =  \left |\bracket{\bar M (t)}{M} \right |^2 = \left | \frac{p}{q}\, g_- (t) \right |^2; \nonumber \\
&& w( M (0) \to M (t)) =  \left |\bracket{M (t)}{M} \right |^2 =  |g_+ (t)|^2; \nonumber \\
&& w(\bar M (0) \to M (t)) =  \left |\bracket{M (t)}{\bar M} \right |^2 = \left | \frac{q}{p}\, g_- (t) \right |^2; \nonumber \\
&&w(M_1^{(2)},\, \bar M^{(1)},\, t) = \left |\bra{M_1^{(2)}}\bracket{\bar M^{(1)}}{\Psi (t)}\right |^2\, =\,\frac{1}{4}\, e^{-2 \Gamma\, t}; \nonumber \\
&&w(M_1^{(2)},\, M^{(1)},\, t) = w(M_2^{(2)},\, \bar M^{(1)},\, t) = w(M_2^{(2)},\, M^{(1)},\, t) =  \,\frac{1}{4}\, e^{-2 \Gamma\, t}. \nonumber
\end{eqnarray}

\section{Functions $\mathrm{R_N}$}
\label{sec:B}

The functions $\mathrm{R_N}(x,\, r,\, \zeta,\, \lambda)$ depend on the
dimensionless parameters $x = \Delta\Gamma t$, $\lambda =
\nicefrac{\Delta M}{\Delta\Gamma}$, the absolute value $r$, and the
phase $\zeta$
of $\nicefrac{q}{p}$. In the most general way these functions can be
written as: 
$$
\label{R_Ncommon}
\mathrm{R_ N} (x,\, r,\, \zeta,\, \lambda) = \\
                  f^{(N)}_0 
          \;+\; f^{(N)}_{0c}\,\cos(2 \lambda x)  
          \;+\; f^{(N)}_{0s}\,\sin(2\lambda x ) 
	 \;+\; f^{(N)}_{1s}\,\sh(x )  
          \;+\; f^{(N)}_{1c}\,\ch(x ) \;+ $$  $$ 
	 \;+\; f^{(N)}_{2sc}\,\sh \left (\frac{x}{2} \right ) \cos( \lambda x )
	 \;+\; f^{(N)}_{2cc}\,\ch \left (\frac{x }{2} \right ) \cos( \lambda x )
	 \;+\; f^{(N)}_{2cs}\,\ch \left (\frac{x}{2} \right ) \sin( \lambda x ),
$$
where we explicitly show the dependence of $\mathrm{R_ N}$ on $x$ and
$\lambda$. The coefficients $f^{(N)}(r,\,\zeta)$ are:

for $N=5$
\begin{eqnarray}
&& f^{(5)}_0(r,\zeta) = 
\frac{5 r^6-4 r^5 \cos (\zeta )+7 r^4-2 r^3 \cos (\zeta )-2 r^3 \cos \
(3 \zeta )+3 r^2+1}{8 r^4 \left(r^2+1\right)}; 
\nonumber \\
&& f^{(5)}_{0c}(r,\zeta) =
-\frac{(r-1) (r+1) \left(r^6-r^4+2 r^3 \cos (\zeta )-2 r^3 \cos (3 \
\zeta )-r^2+1\right)}{16 r^4 \left(r^2+1\right)}; 
\nonumber \\
&& f^{(5)}_{0s}(r,\zeta) =
-\frac{(r-1)^2 (r+1)^2 \sin (\zeta ) \left(r^2-2 r \cos (\zeta )+1\right)}{8 r^3 \left(r^2+1\right)};  
\nonumber \\
&& f^{(5)}_{1s}(r,\zeta) =
\frac{\cos (\zeta ) \left(r^4-2 r^3 \cos (\zeta )+6 r^2-2 r \cos \
(\zeta )+1\right)}{8 r^3};  
\nonumber \\
&& f^{(5)}_{1c}(r,\zeta) =
\frac{r^6+7 r^4-6 r^3 \cos (\zeta )-2 r^3 \cos (3 \zeta )+7 r^2+1}{16 \
r^4};  
\nonumber \\
&& f^{(5)}_{2sc}(r,\zeta) =
\frac{(r-1) (r+1) \cos (\zeta ) \left(r^2-2 r \cos (\zeta )+1\right)}{4 r^3};  
\nonumber \\
&& f^{(5)}_{2cc}(r,\zeta) =
-\frac{r^6+4 r^5 \cos (\zeta )-5 r^4-2 r^3 \cos (\zeta )-2 r^3 \cos \
(3 \zeta )+3 r^2+1}{4 r^4 \left(r^2+1\right)};  
\nonumber \\
&& f^{(5)}_{2cs}(r,\zeta) =
-\frac{(r-1) (r+1) \sin (\zeta ) \left(r^4-2 r^3 \cos (\zeta )+6 \
r^2-2 r \cos (\zeta )+1\right)}{4 r^3 \left(r^2+1\right)};
\nonumber
\end{eqnarray}

for $N=6$
\begin{eqnarray}
&& f^{(6)}_0(r,\zeta) = 
\frac{r^6+3 r^4-2 r^3 \cos (\zeta )-2 r^3 \cos (3 \zeta )+7 r^2-4 r \
\cos (\zeta )+5}{8 \left(r^2+1\right)}; 
\nonumber \\
&& f^{(6)}_{0c}(r,\zeta) =
\frac{(r-1) (r+1) \left(r^6-r^4+2 r^3 \cos (\zeta )-2 r^3 \cos (3 \
\zeta )-r^2+1\right)}{16 r^2 \left(r^2+1\right)}; 
\nonumber \\
&& f^{(6)}_{0s}(r,\zeta) =
\frac{(r-1)^2 (r+1)^2 \sin (\zeta ) \left(r^2-2 r \cos (\zeta )+1\right)}{8 r \left(r^2+1\right)};  
\nonumber \\
&& f^{(6)}_{1s}(r,\zeta) =
\frac{\cos (\zeta ) \left(r^4-2 r^3 \cos (\zeta )+6 r^2-2 r \cos \
(\zeta )+1\right)}{8 r};  
\nonumber \\
&& f^{(6)}_{1c}(r,\zeta) =
\frac{r^6+7 r^4-6 r^3 \cos (\zeta )-2 r^3 \cos (3 \zeta )+7 r^2+1}{16 \
r^2};  
\nonumber \\
&& f^{(6)}_{2sc}(r,\zeta) =
-\frac{(r-1) (r+1) \cos (\zeta ) \left(r^2-2 r \cos (\zeta )+1\right)}{4 r};  
\nonumber \\
&& f^{(6)}_{2cc}(r,\zeta) =
-\frac{r^6+3 r^4-2 r^3 \cos (\zeta )-2 r^3 \cos (3 \zeta )-5 r^2+4 r \
\cos (\zeta )+1}{4 \left(r^2+1\right)};  
\nonumber \\
&& f^{(6)}_{2cs}(r,\zeta) =
-\frac{(r-1) (r+1) \sin (\zeta ) \left(r^4-2 r^3 \cos (\zeta )+6 \
r^2-2 r \cos (\zeta )+1\right)}{4 r \left(r^2+1\right)}; 
\nonumber
\end{eqnarray}

for $N=7$
\begin{eqnarray}
&& f^{(7)}_0(r,\zeta) = 
\frac{5 r^6+4 r^5 \cos (\zeta )+7 r^4+2 r^3 \cos (\zeta )+2 r^3 \cos \
(3 \zeta )+3 r^2+1}{8 r^4 \left(r^2+1\right)}; 
\nonumber \\
&& f^{(7)}_{0c}(r,\zeta) =
-\frac{(r-1) (r+1) \left(r^6-r^4-2 r^3 \cos (\zeta )+2 r^3 \cos (3 \
\zeta )-r^2+1\right)}{16 r^4 \left(r^2+1\right)}; 
\nonumber \\
&& f^{(7)}_{0s}(r,\zeta) =
\frac{(r-1)^2 (r+1)^2 \sin (\zeta ) \left(r^2+2 r \cos (\zeta )+1\right)}{8 r^3 \left(r^2+1\right)};  
\nonumber \\
&& f^{(7)}_{1s}(r,\zeta) =
-\frac{\cos (\zeta ) \left(r^4+2 r^3 \cos (\zeta )+6 r^2+2 r \cos \
(\zeta )+1\right)}{8 r^3};  
\nonumber \\
&& f^{(7)}_{1c}(r,\zeta) =
\frac{r^6+7 r^4+6 r^3 \cos (\zeta )+2 r^3 \cos (3 \zeta )+7 r^2+1}{16 \
r^4};  
\nonumber \\
&& f^{(7)}_{2sc}(r,\zeta) =
-\frac{(r-1) (r+1) \cos (\zeta ) \left(r^2+2 r \cos (\zeta )+1\right)}{4 r^3};  
\nonumber \\
&& f^{(7)}_{2cc}(r,\zeta) =
-\frac{r^6-4 r^5 \cos (\zeta )-5 r^4+2 r^3 \cos (\zeta )+2 r^3 \cos \
(3 \zeta )+3 r^2+1}{4 r^4 \left(r^2+1\right)} 
\nonumber \\
&& f^{(7)}_{2cs}(r,\zeta) =
\frac{(r-1) (r+1) \sin (\zeta ) \left(r^4+2 r^3 \cos (\zeta )+6 r^2+2 \
r \cos (\zeta )+1\right)}{4 r^3 \left(r^2+1\right)}. 
\nonumber
\end{eqnarray}

for $N=8$
\begin{eqnarray}
&& f^{(8)}_0(r,\zeta) = 
\frac{r^6+3 r^4+2 r^3 \cos (\zeta )+2 r^3 \cos (3 \zeta )+7 r^2+4 r \
\cos (\zeta )+5}{8 \left(r^2+1\right)}; 
\nonumber \\
&& f^{(8)}_{0c}(r,\zeta) =
\frac{(r-1) (r+1) \left(r^6-r^4-2 r^3 \cos (\zeta )+2 r^3 \cos (3 \
\zeta )-r^2+1\right)}{16 r^2 \left(r^2+1\right)}; 
\nonumber \\
&& f^{(8)}_{0s}(r,\zeta) =
-\frac{(r-1)^2 (r+1)^2 \sin (\zeta ) \left(r^2+2 r \cos (\zeta )+1\right)}{8 r \left(r^2+1\right)};  
\nonumber \\
&& f^{(8)}_{1s}(r,\zeta) =
-\frac{\cos (\zeta ) \left(r^4+2 r^3 \cos (\zeta )+6 r^2+2 r \cos \
(\zeta )+1\right)}{8 r};  
\nonumber \\
&& f^{(8)}_{1c}(r,\zeta) =
\frac{r^6+7 r^4+6 r^3 \cos (\zeta )+2 r^3 \cos (3 \zeta )+7 r^2+1}{16 \
r^2};  
\nonumber \\
&& f^{(8)}_{2sc}(r,\zeta) =
\frac{(r-1) (r+1) \cos (\zeta ) \left(r^2+2 r \cos (\zeta )+1\right)}{4 r};  
\nonumber \\
&& f^{(8)}_{2cc}(r,\zeta) =
-\frac{r^6+3 r^4+2 r^3 \cos (\zeta )+2 r^3 \cos (3 \zeta )-5 r^2-4 r \
\cos (\zeta )+1}{4 \left(r^2+1\right)};  
\nonumber \\
&& f^{(8)}_{2cs}(r,\zeta) =
\frac{(r-1) (r+1) \sin (\zeta ) \left(r^4+2 r^3 \cos (\zeta )+6 r^2+2 \
r \cos (\zeta )+1\right)}{4 r \left(r^2+1\right)}. 
\nonumber
\end{eqnarray}

\end{document}